\documentclass[fleqn,10pt]{wlscirep}

\usepackage{subcaption}
\newcommand{\rev}[1]{{\color{black} #1}}

\title{Revealing semantic and emotional structure of suicide notes with cognitive network science}

\author[1,2,3,+]{Andreia Sofia Teixeira}
\author[4,+]{Szymon Talaga}
\author[5]{Trevor James Swanson}
\author[6,7,*]{Massimo Stella}

\affil[1]{INESC-ID, R. Alves Redol 9, 1000-029 Lisboa, Portugal}
\affil[2]{Indiana Network Science Institute, Indiana University, 1001 IN-45 Bloomington IN, USA}
\affil[3]{Hospital da Luz Learning Health, Luz Saúde, Avenida Lusíada, 100, Edifício C, 1500-650 Lisboa, Portugal}
\affil[4]{University of Warsaw, The Robert Zajonc Institute for Social Studies, Stawki 5/7, Warszawa 00-183, Poland}
\affil[5]{University of Kansas, Department of Psychology, 1415 Jayhawk Blvd, Lawrence KS, 66045, USA}
\affil[6]{CogNosco Lab, Department of Computer Science, University of Exeter, Exeter EX4 4PY, UK}
\affil[7]{Complex Science Consulting, Via Amilcare Foscarini 2, 73100 Lecce, Italy}
\affil[+]{Andreia Sofia Teixeira contributed equally to this work with Szymon Talaga.}
\affil[*]{Corresponding author: massimo.stella@inbox.com}


\begin{abstract}
Understanding how people who commit suicide perceive their cognitive states and emotions represents an important open scientific challenge. We build upon cognitive network science, psycholinguistics and semantic frame theory to introduce a network representation of suicidal ideation as expressed in multiple suicide notes. By reconstructing the knowledge structure of such notes, we reveal interconnections between the ideas and emotional states of people who committed suicide through an analysis of emotional balance motivated by structural balance theory, semantic prominence and emotional profiling.
Our results indicate that connections between positively- and negatively-valenced terms give rise to a degree of balance that is significantly higher than in a null model where the affective structure is randomized and in a linguistic baseline model capturing mind-wandering in absence of suicidal ideation. We show that suicide notes are affectively compartmentalized such that positive concepts tend to cluster together and dominate the overall network structure. Notably, this positive clustering diverges from perceptions of self, which are found to be dominated by negative, sad conceptual associations in analyses based on subject-verb-object relationships and emotional profiling.
A key positive concept is ``love'', which integrates information relating the self to others and is semantically prominent across suicide notes. The emotions constituting the semantic frame of ``love'' combine joy and trust with anticipation and sadness, which can be linked to psychological theories of meaning-making as well as narrative psychology. Our results open new ways for understanding the structure of genuine suicide notes and may be used to inform future research on suicide prevention.
\end{abstract}

\begin{document}

\flushbottom
\maketitle
\thispagestyle{empty}

\subsection*{Introduction}

In many circumstances, the only difference between a completed suicide and a suicide attempt is slightly greater pressure applied to a trigger. In either case the importance of gaining a greater understanding of the psychological conditions surrounding such a tragic event is immediately apparent; what leads an individual to contemplate, and perhaps commit, such an act? Similarly, what sorts of thoughts and feelings does one encounter when experiencing suicidal ideation? And how might our understanding of these phenomena aid in improving prevention efforts? Although these are challenging questions, suicide notes represent one potential window into the psychology of individuals who complete suicide~\cite{mcadams2001psychology,proulx2006death}. By analyzing the language and contents of suicide notes, we can gain unique insight into shared features of individual experiences and perhaps a greater understanding of the cognitive processes that accompany suicidal ideation~\cite{schneidman1981suicide}.

Previous research on suicide notes has highlighted specific properties of such notes in an attempt to better understand what characteristics stand out and differentiate them from other types of texts. Some work has focused on studying the contents of suicide notes~\cite{foster2003suicide,handelman2007content,o1999thematic}, including dominant emotional themes (such as ``anger'' and ``love''), and key motives (e.g., the wish to die). In general, these studies have focused on answering the question: what is in a suicide note? That is, what are the contents that we most consistently observe when comparing notes from people that committed suicide? For example, Al-Mosaiwi and Johnstone~\cite{al2018absolute} recently found that the vocabulary used by individuals at risk of suicide was different from those who suffered from other mental disorders related to depression and anxiety. Individuals who experienced suicidal ideation tended to utilize different vocabularies and mainly absolutist words, indicating that suicide notes have their own emotional and lexical footprints.

Sentiment analysis has been further applied to the goal of comparing how the emotional contents of suicide notes are categorized by learning algorithms versus trained clinicians~\cite{pestian2012sentiment,schoene2016automatic}, as well as whether or not such algorithms can reliably distinguish between genuine and simulated suicide notes~\cite{pestian2010suicide}.
These automated text-analysis techniques offer some powerful advantages over the standard, qualitative approaches that have commonly been applied to the study of suicide notes by clinical psychologists. For one, quantitative methods -- such as those used in sentiment analysis~\cite{pestian2010suicide,schoene2016automatic} and emotional profiling~\cite{mohammad2013crowdsourcing,stella2020text} -- allow for the use of more objective criteria and clearer operationalizations of psychological constructs (such as valence, emotional intensity, and others based on normative data). Qualitative methods, on the other hand, depend upon human judges to code and interpret texts, then compare ratings to assess the consistency of their conclusions. Thus, there may be a high degree of uncertainty and limited reliability when such techniques are used to make inferences. On the other hand, complex statistical/machine learning models often produce results that are difficult to understand and thus may not be very helpful for tasks different than prediction such as explanation~\cite{rudin2019stop}.


In the present work we apply network science methods to the analysis of genuine suicide notes. Importantly, we show how network modeling can be used to expand the text-analytic toolbox in psychology and provide novel ways of answering complex research questions about text data
~\cite{castro2020contributions}.
In contrast with other automated approaches to text analysis~\cite{hassani2020text}, network models allow researchers to encode not only, e.g., word sentiment, but also the broader set of connections that each word has with its surrounding text~\cite{stella2020text}. \rev{This allows one to track not only \textit{which} words appear more or less often in a sample of texts but also \textit{how} they are used and in what contexts, thereby affording comparisons with other linguistic baselines~\cite{de2019small}.} Additionally, unlike typical black box models~\cite{rudin2019stop}, network methods are fully transparent and produce results which are often much easier to interpret. \rev{By mapping the interrelationships among words in accordance with their usage, networks offer a clearer window into semantic frames which endow words with their broader meaning. This explicit mapping helps to both avoid the oversimplification of information conveyed in written texts as well as facilitate understanding. Additionally, network models open a window into the meaning of words by explicating meaningful semantic or syntactic relationships between concepts. Network representations make it possible to observe and understand the usage of words in terms of associations with other ideas, creating a “map” of knowledge that reflects the ways text authors think and can then be reconstructed by researchers~\cite{lee2015coding}.
} Hence, network modeling represents an approach to the study of text data that can further elucidate the structure of human texts~\cite{correa2020semantic} and potentially reveal how concepts are perceived, organized and interconnected in the human mind~\cite{castro2020contributions}.

\begin{figure}[t!]
    \centering
    \includegraphics[width=.48\textwidth]{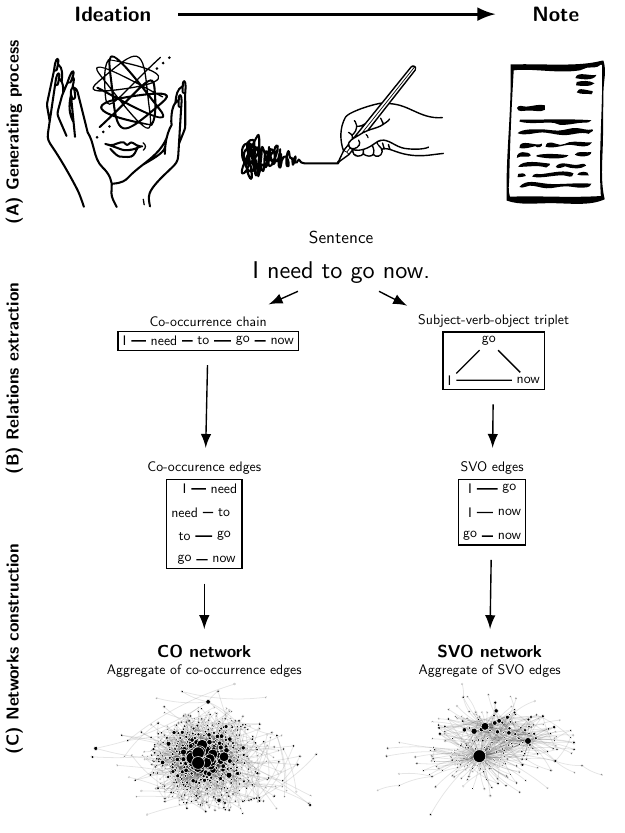}
    \caption{
        Schematic depiction of the data generating process (A), the method of relations extraction (B) and the network construction scheme (C). As (A) suggests, the assumption of our approach is that suicide notes are observable, even if noisy, realizations of unobservable suicidal ideation processes.
    }
    \label{fig:method-diagram}
\end{figure}

\subsection*{The relevance of networks in understanding suicide notes}

Language guarantees an expression of people's perceptions through semantic content and emotions. Semantic frame theory~\cite{fillmore2006frame} indicates that the meaning attributed by people to a given concept can be reconstructed by observing the relationships and conceptual associations attributed to that concept in text or speech. Words in a given semantic frame elicit different combinations of emotions, i.e. emotional profiles, which characterize the emotional content of a text.

Network science provides tools for quantifying and reconstructing both semantic frames~\cite{fillmore2006frame,ribeiro2019semantic,correa2020semantic} and emotional associations~\cite{stella2020text}, serving as a framework for the quantitative identification of ways in which people perceive events and happenings~\cite{fillmore2006frame,aitchison2012words,plutchik2013theories,castro2020contributions,stella2020text}. In comparison to more opaque machine learning techniques, networks have the advantage of transparently representing a proxy for the associative structure of language in the human mind, within the cognitive system apt at acquiring, storing and producing language, i.e. the mental lexicon~\cite{aitchison2012words,vitevitch2019network}. Supported by psycholinguistic inquiries into the mental lexicon~\cite{vitevitch2019network,kenett2017semantic,stella2017multiplex,castro2020contributions}, complex networks built from texts can open a window into people's mindsets~\cite{stella2020text}. Focus here is given to reconstructing the collective mindset as expressed in the last written words left by people who committed suicide. \rev{To this aim, we adopt a corpus of genuine suicide notes gathered in a previous study~\cite{schoene2016automatic} and including 139 letters from people who committed suicide. The letters come from a collection of suicide notes from sources like newspapers, books and diaries collected by clinical psychologists mostly in the US and in Europe. The notes were written and collected over a time window spanning over 60 years and have been used also for recent machine learning approaches to automatic detection of suicide ideation \cite{schoene2016automatic}. On average, a letter included 120 $\pm$ 12 words and a total of over 2000 different concepts were stated in the whole corpus.}

\subsection*{Cognitive network approach to suicidal ideation analysis}

In this manuscript we consider the content of suicide notes an observable realization of otherwise unobservable mental states and suicidal ideation of their authors~\cite{vitevitch2019network}. In order to map out relationships between main concepts and the emergent semantics of suicide notes we reduced raw texts to two different network representations: co-occurrence (CO) and subject-verb-object (SVO) networks. \rev{Co-occurrence networks capture mostly successive relationships between adjacent concepts in a sentence while SVO networks capture syntactic links between actors and actions as identified with natural language processing and described in suicide letters.} See Fig.~\ref{fig:method-diagram} and Materials and Methods section for details.

\rev{Unlike several well-known studies of semantic
networks~\cite{steyvers2005large,teixeira2010complex}
based on semantic associations stored in lexical databases,
in our approach networks of associations are extracted directly from raw texts
as written down by individual people. As such it can be seen as an extension
of map analysis, as used by Carley~\cite{carley1986approach,carley1997extracting},
enriched with: (i) link extraction and analysis based on modern natural language processing and network science metrics, (ii) additional cognitive data about affect patterns used in synergy with network structure, and (iii) linguistic benchmarks relying on recent datasets of conceptual associations (namely, free associations, see next section). These three points represent key ways in which we build upon and extend previous methods.}

\rev{Notice that our ultimate goal is to identify mental and conceptual associations which are on average most typical for suicide notes. Hence, we do not take a sociological perspective focused on evolution of collective narrative strategies in connection with other social processes~\cite{padgett2020political,fuhse2020relating}. Instead, we aim at identifying cognitive patterns which are common across different individuals who committed suicide.}

\subsection*{\rev{Using free associations as linguistic benchmarks for text analysis}}

Unlike previous approaches, we do not aim to discriminate suicide notes from other types of text~\cite{pestian2012sentiment,schoene2016automatic}. Instead, we focus on a quantitative understanding of the mindsets of people who committed suicide. \rev{To this aim, it is important to compare the structure of conceptual associations found in suicide notes against a linguistic benchmark. Since the focus of our work lies on conceptual associations and cognitive networks, we did not select another corpus of text as a benchmark but rather utilized a separate network dataset of free associations (FA)} \cite{steyvers2005large,de2019small}. \rev{These associations capture the structure of semantic memory and indicate how individuals associate concepts with one another during mind-wandering, i.e. while thinking freely without additional semantic, phonological or syntactic constraints} ~\cite{de2019small}. \rev{This aspect of concept recollection makes free associations particularly appealing for investigating how the flow of specific narratives differs from mind-wandering. For this reason, we use the FA network, as obtained from the "Small-World of Words" project \cite{de2019small}, as a linguistic benchmark for comparison against suicide notes.}

\subsection*{Manuscript structure}

Study 1 investigates the ``emotional syntax'' of suicide notes, analyzing whether the connectivity and configuration of words is somehow related to their valence. We use \textit{structural balance theory}~\cite{heider} to assess the degree of balance in the network and determine how valence is organized among neighboring words. We extend previous research by: (i) studying the emotional content of suicide notes and (ii) mapping how sentiment is organized in the collective mindset around suicidal ideation. Study 2 focuses on subject-verb-object relationships to highlight self-perceptions in suicide notes. Study 3 combines network centrality, semantic frames and emotional data in order to describe and quantify typical emotions associated with different concepts in suicide notes. We conclude with a general discussion of the relevance of this study \textit{vis-à-vis} previous results and current gaps in the literature.

\section*{Study 1: Investigating Structural Balance in Suicide Nodes Co-occurrence Networks}

To assess the degree of balance of the CO network we assign signs to edges based on the presence of negative and positive sentiment labels of words (see Figure~\ref{fig:signednet} and Materials and Methods section for more details).


\subsection*{\rev{Triadic closure, emotional balance and its interpretation in suicide notes}}

Based on sentiment labels of words, we introduce \rev{an analysis inspired by} structural balance. Structural balance theory has its origin in a study by Fritz Heider, in 1946, that evaluated the psychological and cognitive configurations of interpersonal relations positioned in a triad~\cite{heider}. These relations can be positive or negative and include feelings such as friendship, love, esteem, as well as their opposites. Heider stated that for a triad to be balanced it must have an even number of negative relations, otherwise tension emerges. Structural balance has also been adapted and used to study different complex systems represented as signed networks: from adaptive behavior in social networks~\cite{teixeira2017emergence,he2018evolution} to financial networks~\cite{souto2018capturing,harary2002signed}, among others~\cite{aref2019balance,doreian2015structural,leskovec2010predicting}. \rev{This characterization of triadic closure has also been investigated in terms of psychological states and clinical conditions \cite{moradimanesh_altered_2021,chiang2020triadic}. In fact, structural balance theory has been recently used to detect patterns of tension or stress in signed networks coming from cognitive neuroscience. In a study concerned with stress in the context of cooperative dilemmas, subjects engaging in unbalanced triads activated brain regions concerned with processing distress and cognitive dissonance more frequently~\cite{chiang2020triadic}. Another recent approach shifted this psychological connotation of structural balance from social ties to the way brain regions activate or get inhibited over time in clinical populations. By analysing brain functional connectivity, Moradimanesh and colleagues showed that there were over-represented balanced triads in the brain network of people diagnosed with an autism spectrum disorder in comparison to healthy controls~\cite{moradimanesh_altered_2021}. Notice that in the context of cognitive neuroscience, the term "balance" does not imply positive social interactions but rather positively signed brain activity (e.g.~positive correlations between firings of different regions). Along this view, the increased number of balanced triads reported by \cite{moradimanesh_altered_2021} reflects an increased brain activity in people diagnosed with an autism spectrum disorder. }

\rev{In the present study, we follow the above interpretations by extending the meaning of "balance" to encompass properties of affective organization within the context of signed networks built from conceptual associations. As such we refer to this extended interpretation as \textit{emotional balance.} Specifically, we study how} co-occurrence relationships and valences (sentiments) of words in suicide notes \rev{can provide information about possible psychological tensions, distress and emotional disturbances \cite{chiang2020triadic} within narrative structures}. \rev{To yield positive and negative links between words, we adopt the following reasoning:}
\rev{\begin{itemize}
    \item If there is a negative word in a triad, the links connected to it will also be negative, representing emotional tension between positive and negative concepts;
    \item If there are two negative words, we assume the link is still negative given that the conceptual association is bridging ideas perceived as negative and thus creating additional tension;
    \item Links between positive words will be positive;
    \item If there is a neutral word in the triad, the link will retain the valence of the other end node possessing a sentiment polarity;
    \item If there are two or more neutral words in the triad, we do not consider the triad as being either balanced or unbalanced.
\end{itemize}
}

\begin{figure}[h!]
\centering
\begin{subfigure}[b]{0.48\textwidth}
\centering
\includegraphics[width=0.8\textwidth]{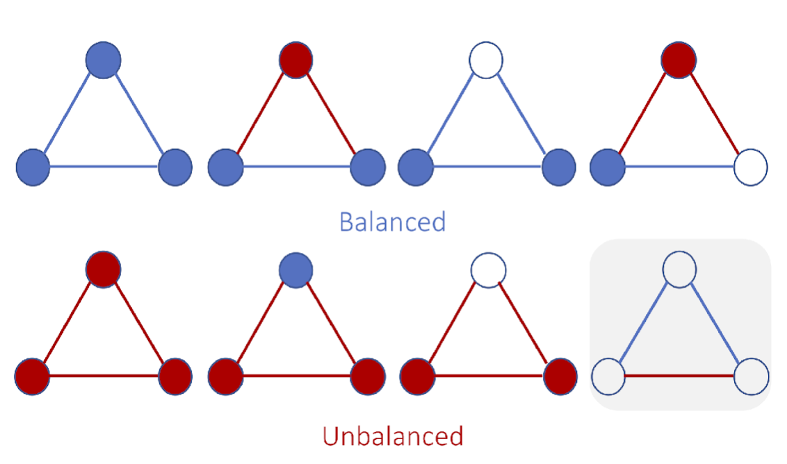}
\caption{Signed triads.}\label{fig:signednet}
\end{subfigure}
\begin{subfigure}[b]{0.46\textwidth}
\centering
\hfil
                \raisebox{.12\textwidth}{
\includegraphics[width=\textwidth]{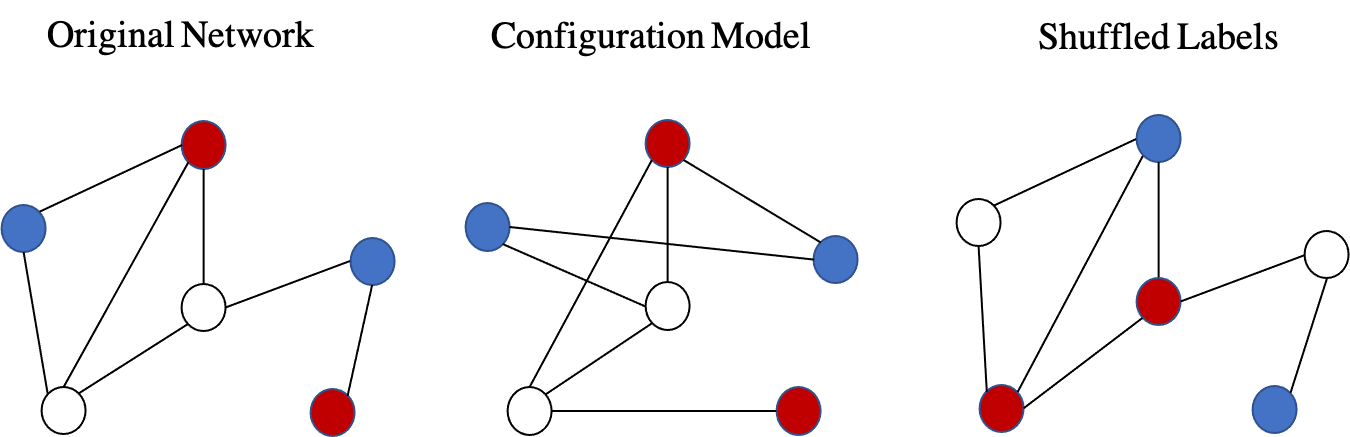}
}
\caption{Illustration of the null models.}\label{fig:nullmodels}
\end{subfigure}

\centering
\caption*{\textbf{Frequency of triads}}
\begin{subfigure}[b]{0.47\textwidth}
\includegraphics[width=\textwidth]{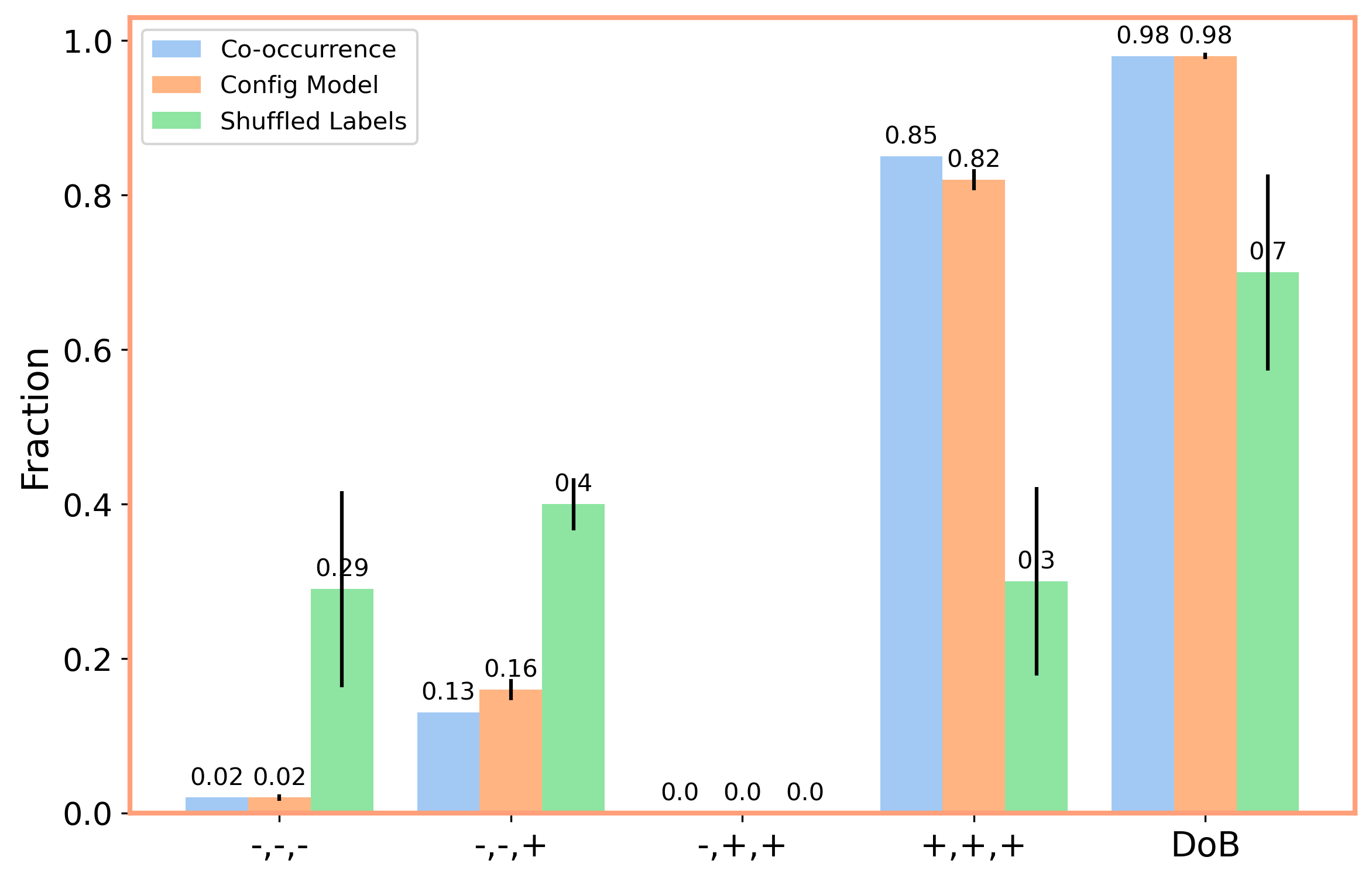}
\caption{CO Network}\label{sb:co}
\end{subfigure}
\hfil
\begin{subfigure}[b]{0.47\textwidth}
\includegraphics[width=\textwidth]{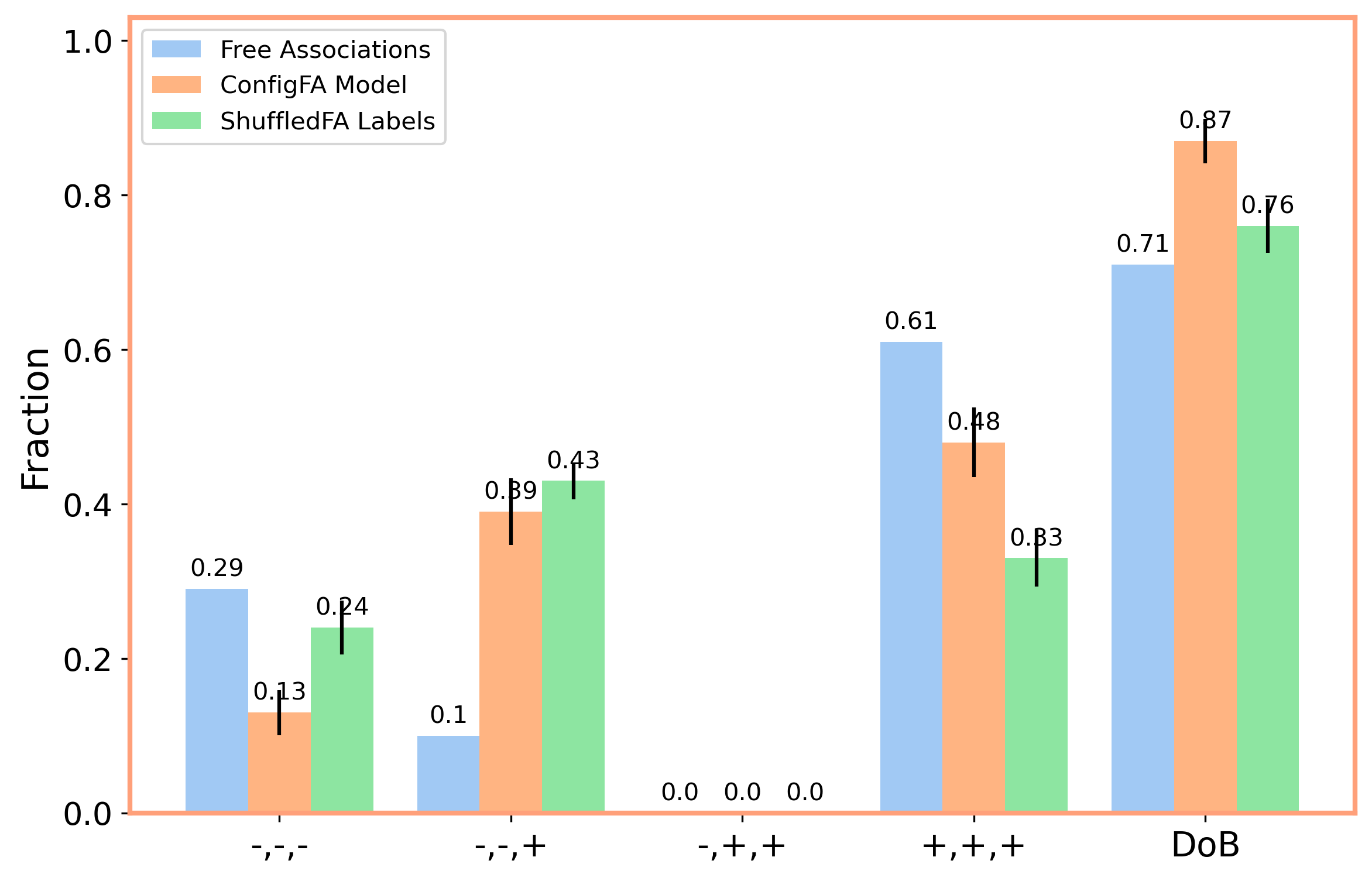}
\caption{FA Network}\label{sb:fa}
\end{subfigure}
\caption{On panel \textbf{(a)} we represent balanced and unbalanced triads along with the valence labels that generate those triads. \rev{Blue color represents positive concepts/links, red color represents negative concepts, negative links, and white circles represent neutral concepts. As explained in the text, t}he shadowed triad is the only unbalanced configuration impossible to obtain given any combination of positive, negative and neutral words. On panel \textbf{(b)} we illustrate the two null models used. Starting with a toy network we illustrate the process of rewiring links, while keeping the degree distribution with the Configuration Model, and then of shuffling the sentiment labels, while keeping the structure of the original network. On panel \textbf{(c)} we show the fraction of each triad and the total degree of balance for the CO Network and its correspondent null models and on panel \textbf{(d)} we present the same statistics for the FA network. When presenting the null models we provide the average and the standard deviation over 1000 realizations.}\label{fig:sb}

\end{figure}

\rev{Given this setting, in Figure~\ref{fig:signednet} we show how we obtain emotionally balanced and unbalanced triads based on the valence of the words. We consider a triad emotionally balanced if there are a majority of positive words or if there is no predominant valance.}

\rev{If negative concepts are predominant, we consider the triad unbalanced. In this sense, emotional unbalance reflects the preponderance of negatively-valenced words within a given triad. Our definition of emotional balance corresponds to the definition of strong structural balance considering $\{-, -, -\}$ triads as unbalanced \cite{cartwright1956structural}.}

\rev{Following this definition, a triad with a negative word will become balanced if the other two end nodes are positive. An all-positive triad emerges from all positive words or from two positive words with a neutral word. Neutral triads, since they do not contribute to a positive/negative predominance of affect, are not considered (see Figure~\ref{fig:signednet}). This emotional construction aligns with the definition of balanced ($\{+, +, +\}$ and $\{+, -, -\}$) and unbalanced ($\{-, -, -\}$) triads in structural balance \cite{cartwright1956structural}, with the exception that in our application the formation of the unbalanced triad $\{+, +, -\}$ is not possible.}

\rev{To study the emotional balance structure of mental states in suicide notes we} start by building a signed network from the CO network (see Materials and Methods section); then we evaluate its \rev{triad frequency and} degree of balance -- i.e., fraction of balanced triads; finally we present their statistical significance by comparing observed data against two different null models, as well as against the FA network, proceeding similarly as with the CO network.
\rev{Let us underline that the comparison between the CO network, coming from suicide notes, and the FA network, coming from mind-wandering in the absence of suicide ideation, aims to identify potential patterns of psychological distress expressed in the language of suicide notes and potentially missing from the mental states of healthy controls during free mind-wandering.}

\subsection*{\rev{Emotional balance and triad significance} in suicide notes}

We evaluate \rev{triad frequency} in the signed CO and FA networks and evaluate their statistical significance when compared with two null models. As one model, we use configuration models~\cite{viger2005fast} to generate random networks \rev{where words had the same network degree} as in the CO/FA networks. We also generate random networks using the same exact structure of the CO/FA networks but shuffling the sentiment labels associated with each node.
This second null model does not change the overall structure of the network but instead changes local properties (i.e. degrees) of positive, neutral and negative words relative to their characteristics in the empirical network.
Notice that the signs of edges in each triad are determined by the valence of the labels associated with the nodes so different sentiment label reshufflings induce different edge signs.

We explore how triad frequency and degree of balance (DoB) in both the CO and FA networks differ from random expectation when sequential structure or affective structure is disrupted, respectively. The first comparison is obtained by configuration models randomising network structure but keeping fixed the degree of individual words. The second comparison is obtained by label-shuffling, which altered the location and local degree of words of different valence while keeping the empirical structure fixed. In Figure~\ref{fig:nullmodels} we present a toy example of the null models.

After creating the above modified versions of the networks, we build the associated signed networks and calculate the corresponding degree of balance and triad frequency values for both the null models. We present the mean and standard deviations for each count and for the degree of balance based on a sample of 1000 random realizations.

In Figures~\ref{sb:co} and~\ref{sb:fa} we present the results. We observe that the overall degree of balance and the frequency of $\{+,+,+\}$ triads in the CO network and its configuration models are much higher than for the label-shuffled networks (Fig.~\ref{sb:co}).
In Fig.~\ref{sb:fa} we further provide evidence that the CO network exhibits a high level of degree of balance through comparison with the FA network. In the free association network there is a more uniform distribution of triad frequencies and the null models follow the decrease of all positive triads in comparison to the co-occurrence network of suicide notes.

\subsection*{\rev{Emotional} balance in suicide notes as a narrative strategy}

The results indicate that the \rev{emotional} structure of conceptual associations in suicide notes is more compartmentalized where the all-positive triads ($\{+,+,+\}$) are most frequent. \rev{Mental contents are typically described as compartmentalized when affective information is organized, with respect to a given construct, so that positive and negative associations are relatively segregated and those of the same polarity cluster together.} Compartmentalized mental contents have been investigated within social psychological studies looking at vulnerability and resilience to depression~\cite{showers1992compartmentalization,campbell1991cognitive,showers1998dynamic}. \rev{One difference between the present usage of `compartmentalization' is that previous researchers have focused on the affective organization of the self-concept, whereas we apply this concept to interpreting the entire mental space of authors of suicide letters.} It is possible that the highly balanced pattern we observe in the co-occurrence network is indicative of a psychological strategy for coping with the ``psych-ache'' associated with suicidal ideation (cf.~\cite{schneidman1981suicide,foster2003suicide}). \rev{Namely, this higher degree of organization may suggest an active process of motivated cognition \cite{cramer2016can}. That is, the individual may be driven to associate positive concepts with one another more readily than they would in a free association context (such as mind-wandering in the absence of suicidal ideation) to increase the salience and accessibility of positive-affective information.} This preliminary evidence calls for the need to conduct future survey-based research in order to investigate this possibility.

\section*{Study 2: Perceptions of self via subject-verb-object relationships}

In this study we use the subject-verb-object (SVO) network to identify central actors and in particular the way the self (``I'') is related to other people and concepts in suicide notes. The SVO representation is more suitable for this task than the co-occurrence representation as the number of relations a given word is involved in is not determined by the number of its occurrences alone but rather by the centrality of its position in the syntactic structure of sentences in which it occurs. Hence, in the SVO representation most important sentence-building and therefore meaning-making entities have most central position in the network by design (see~Network~construction section in Supplementary Information (SI) for more details). Therefore, in this approach it is justified to consider node strength \rev{(sum of edge weights)} an appropriate measure of centrality.

\begin{figure}[h!]
\centering
\includegraphics[width=.7\textwidth]{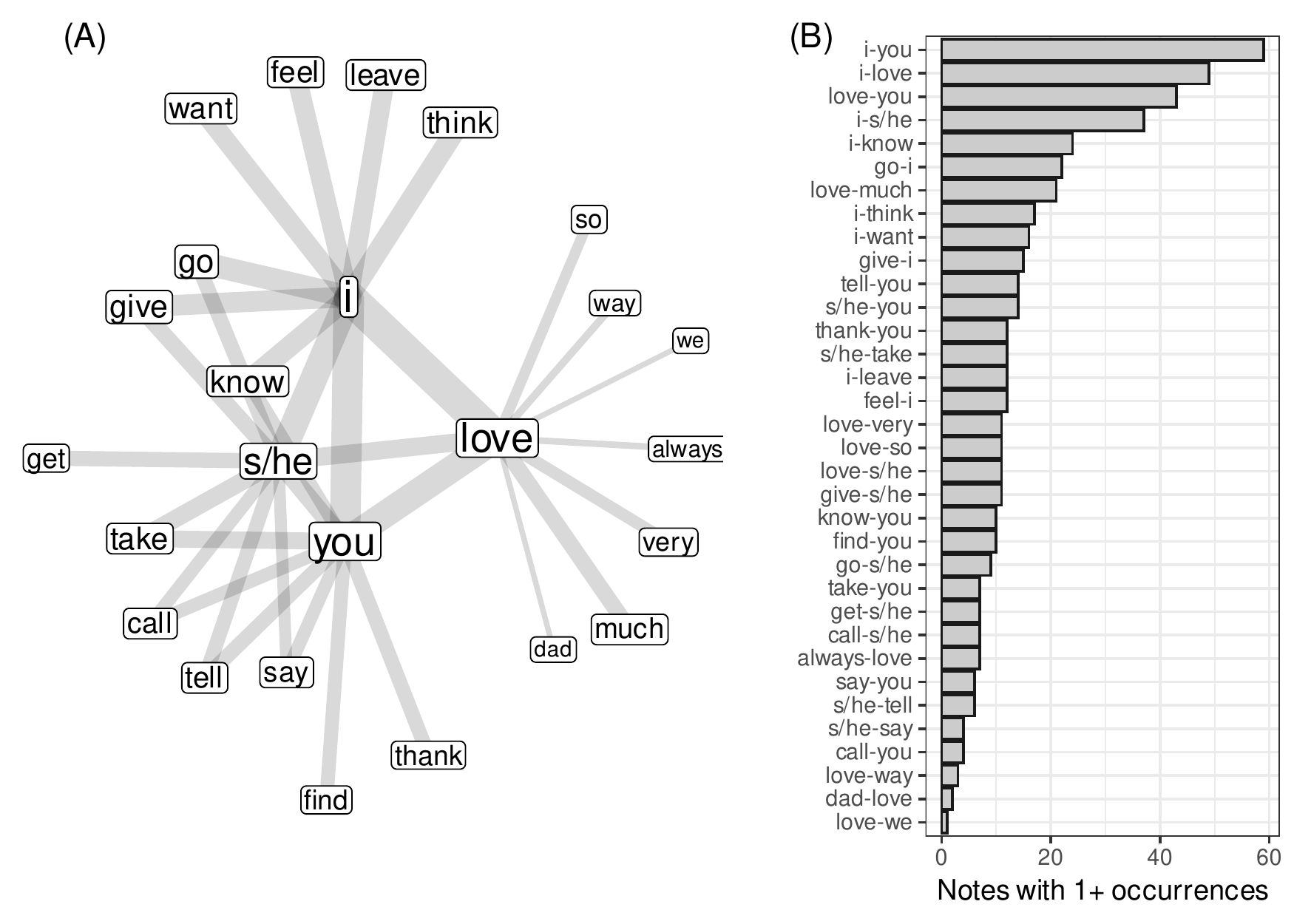}
\caption{
 (A) Network of most frequent relations involving ``i'', ``you'', ``s/he'' and ``love''
 derived from the SVO network.  Arguably, neighbors of the above four words agree with
 what could be expected based the on common sense.
 (B) Number of documents containing specific (unidrected) pairs of concepts. Pairs which may be expected to be typical for suicide notes such as ``i-you'', ``i-love'',
 ``love-you'' and ``go-i'' occupy the top of the plot.
}
\label{fig:svo-pairs}
\end{figure}

\subsection*{Core actors and themes}

\begin{figure}[h!]
\centering
\begin{minipage}{.66\textwidth}
\centering
\includegraphics[width=\textwidth]{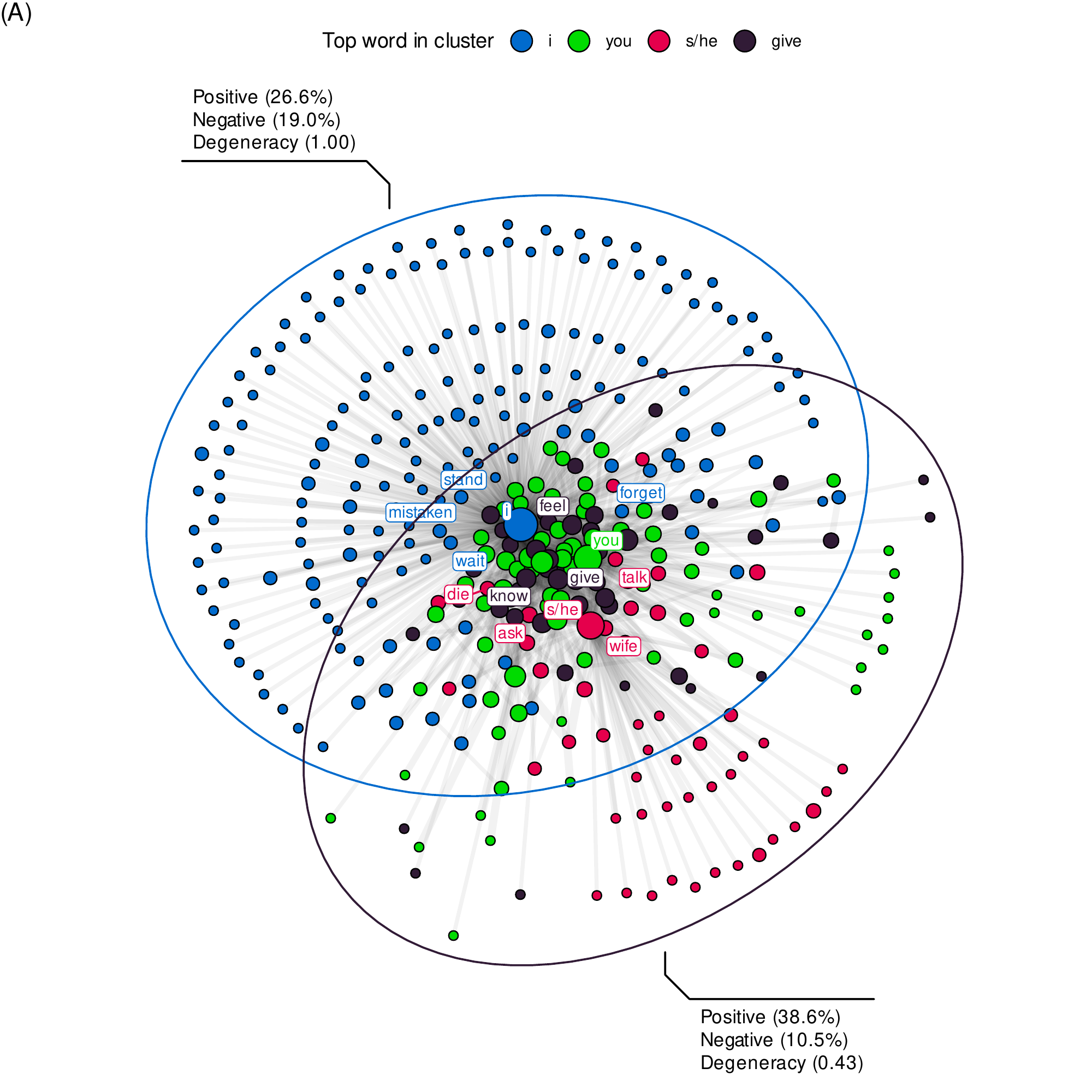}\label{fig:3a}
\end{minipage}
\hfill
\begin{minipage}{.33\textwidth}
\includegraphics[width=\textwidth]{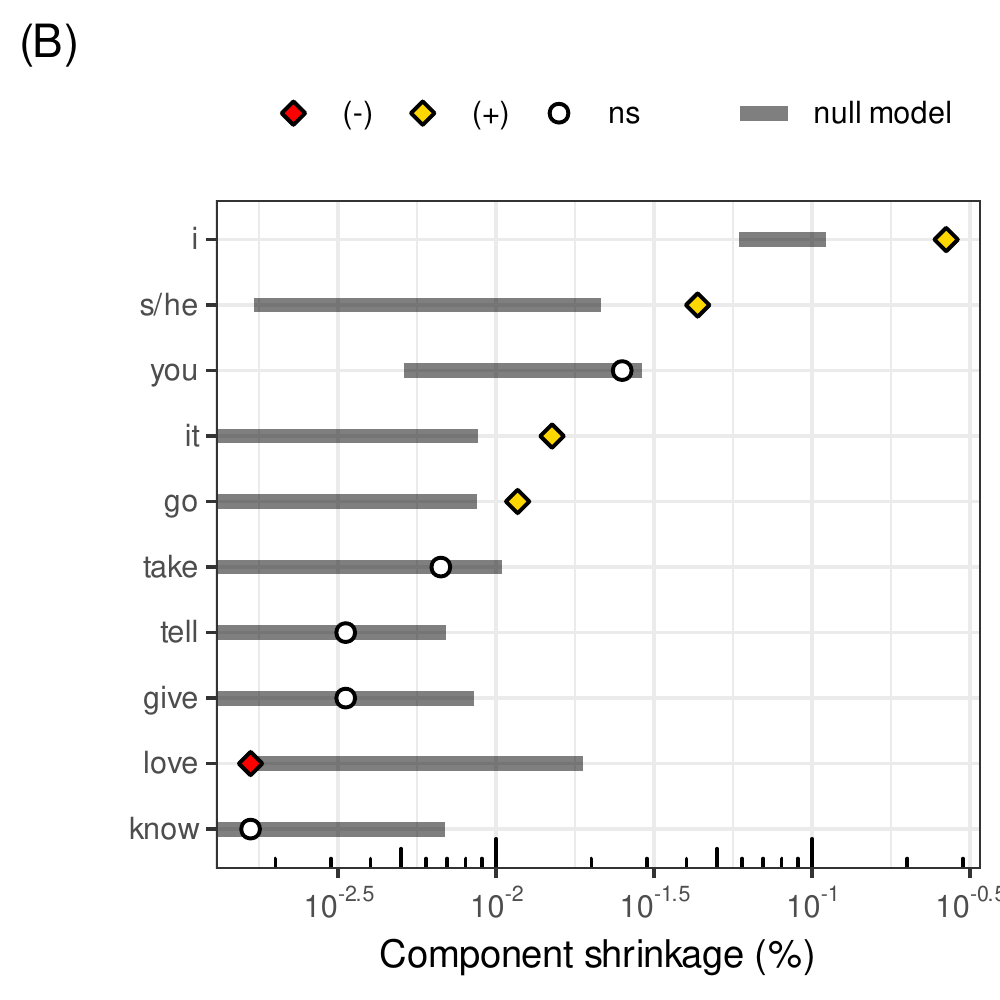}\label{fig:3b}
\includegraphics[width=\textwidth]{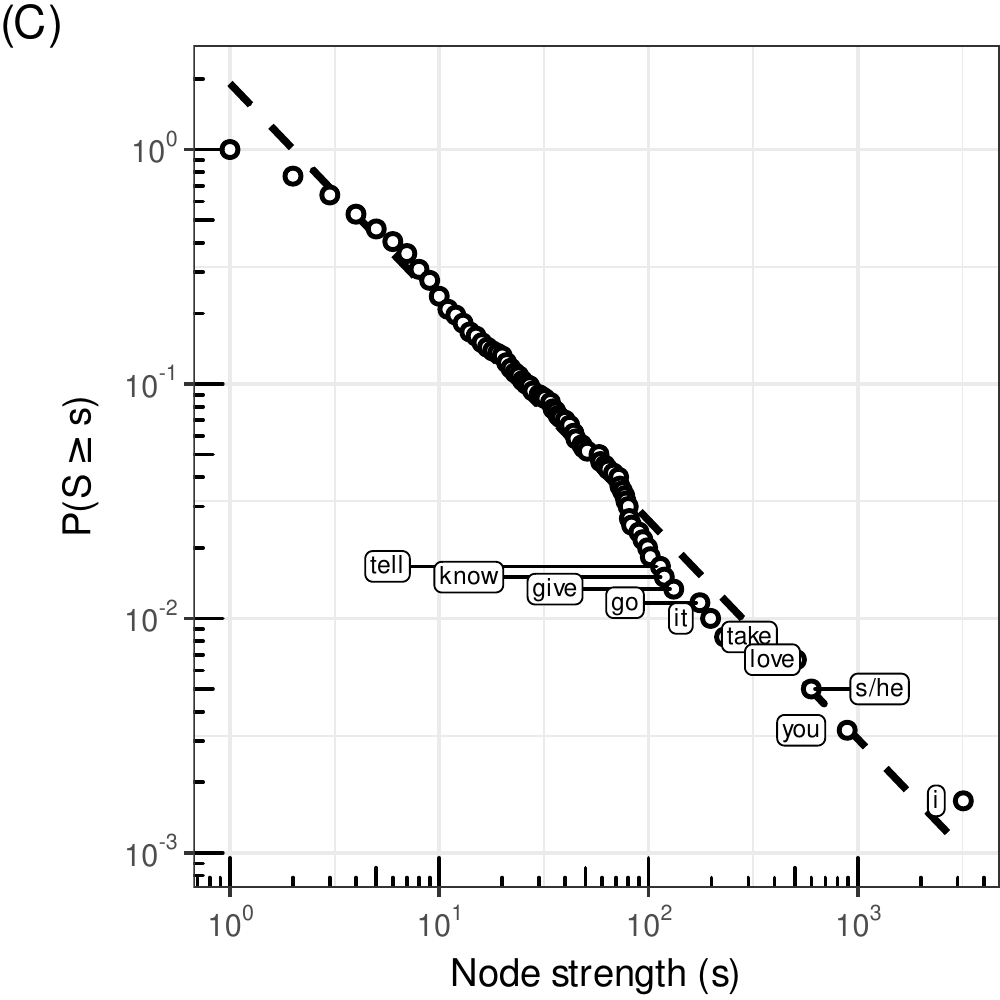}\label{fig:3c}
\end{minipage}
\caption{
    (A)~Four largest clusters in the giant component of the SVO network (containing 89.7\% of nodes and 98.4\% of the total nodes' strength) as detected with G-N algorithm \cite{newman2004finding} (edge weights were not used). The ellipses show general orientations of the largest cluster organized around ``I'' and the remaining clusters. Positive/negative values refer to fractions of words with positive/negative valence. Degeneracy is between 0 and 1 and measures how a (sub)graph is similar to a star graph.
    (B)~Shrinkage factors of the giant component (in log scale) after deleting a node for ten largest hubs relative to a corresponding null model distribution based on 1000 random samples (gray bounds show 1st and 99th centiles). (C) Complementary cumulative distribution function (CCDF) for node strength in log-log scale. The distribution is heavy-tailed (with tail exponent estimated~\cite{clauset2009power} to be $\gamma \approx 2$). As a result, the topology of the network is determined primarily by a few hubs, in particular personal pronouns.
}
\label{fig:svo-main}
\end{figure}

Nodes with highest strengths in clusters (Fig.~\ref{fig:svo-main}A) as well as largest hubs in general (``I'', ``you'', ``s/he'', ``love'', ``take'', ``it'', ``go'', ``give'', ``know'', ``tell''; see~Fig.~\ref{fig:svo-main}C) intuitively seem to be representative of what may be typical messages that people might try to convey in their last written statements. The core actors and themes involved are self and others and relations of love (or lack thereof), taking/giving, going and telling. Figure~\ref{fig:svo-pairs} presents most frequent words and associations in the SVO network as well as the distribution of frequency of most common pairs of words among different suicide notes. Additional validation of the extracted associations is given in Table~3~(SI). Crucially, more insight into typical conceptual associations expressed in suicide notes can be gained by studying the structure of the SVO network in more detail.

As indicated by Figure~\ref{fig:svo-main}C the node strength distribution is heavy-tailed and hence the structure of the SVO network is determined primarily by a few hubs. Interestingly, these are personal pronouns and in particular ``I'', which indicates that self and other persons are main sentence-building and thus meaning-making entities in suicide notes. Clearly, the defining feature of the network is the opposition between a star-like cluster around ``I'' and much less centralized remaining parts of the network. Crucially, it is personal pronouns, such as ``I'', ``you'' and ``s/he'', which play central roles, indicating that relationships with other people (as well as self) are of crucial importance in suicide notes. This seems to reflect their personal character and focus on relations between authors and other people as well as relationships of authors with themselves.

Furthermore, the neighborhood of ``I'' is very star-like and its sentiment/valence is significantly more negative (19\% of words) than the rest of the network (12.2\%), $p = 0.040$, and less positive (26.6\% relative to 36.8\%), $p = 0.020$ ($\chi^2$ test for two proportions). This suggests that perceptions and cognitions about the self are both prevalent and largely isolated from other contexts (words connected to ``I'' often are not linked to any other word) as well as associated with more negative emotions on average. In general, it seems that the defining feature of the network is the division between the star-like part which connects to the rest almost exclusively through ``I'' and the remaining part which is markedly less centralized. We make this argument more rigorous with the notion of degeneracy~\cite{klein_emergence_2020}, which measures the tendency that a random walker starting from a random node after one step ends up in a limited set of central nodes (see~SI:~Network degeneracy) and is normalized between $0$ and $1$, where $1$ is attained only by perfect star graphs (see~Fig.~\ref{fig:svo-main}A).

\subsection*{Degeneracy and component shrinkage in the SVO network}
The degeneracy of the network around ``I'' means that there is a relatively
large subset of 139 nodes (23.2\%) which is linked to the giant component only through ``I'' (see~Fig.~\ref{fig:svo-main}B). It also means that 33.7\% of negative words (29 out of 86) are linked to the network this way. No other node plays as an important role for the overall connectivity and negative sentiment. Moreover, this property is not induced by the strength distribution alone, which implies it is a higher order characteristic indicative of a special role of the self in suicidal ideation.

This phenomenon becomes evident in the comparison of the shrinkage of the giant component after node deletion in the observed SVO network relative to an appropriate null distribution based on undirected weighted (soft) configuration model~\cite{squartini_unbiased_2015} (see~Fig.~\ref{fig:svo-main}B). The null model we use corresponds to a null hypothesis that words connect with each other at random conditional only on their strengths, that is, twice the number of SVO triplets they occur in. Other hubs do not differ as much in terms of their component shrinkage factors (except ``s/he'', but the absolute difference even in this case is much lower) and in general they tend to be either typical representatives of the null model or only slightly positive outliers. The only exception is ``love'' (Fig.~\ref{fig:svo-main}B) which has significantly lower shrinkage factor, $p = 0.007$. In other words ``love'' has both high strength and degree and very few neighbors with no other connections. Hence, ``love'' is rather knitting the whole SVO network together rather than having its own specific set of associations.

\subsection*{SVO structure and self-perception in suicide notes}

In summary, the SVO-based analysis shows that in suicide notes: 1)~the self plays a central role, in particular as the main hub through which negatively-valenced concepts are connected; 2)~the large-scale structure of the concept-network is determined primarily by personal pronouns -- ``I'', ``s/he'' and ``you'' -- corresponding to self and others; 3)~``love'' is the main verb/noun and it does not have a significant unique neighborhood of associated concepts but rather knits together regions focused around the three main pronouns. Qualitatively, this suggests that, on average, a significant feature of suicidal ideation may be a pronounced (perhaps excessive) and largely negative focus on self as well as personal, intimate relations with other people.

This result echoes decades of research in clinical psychology~\cite{beck1976cognitive,ingram1983depression,derry1981schematic} showing that negative self-schemata, as well as enhanced processing of negative self-referential information, play an important role in the emergence and persistence of depressive symptoms. Moreover, considering this finding alongside Study 1, these results are potentially suggestive of the so-called ``hidden vulnerability of compartmentalization''~\cite{zeigler2007self}. That is, while positively-valenced mental contents may be more readily associated with one another, these are not necessarily connected with the self-concept, which may instead be dominated by negatively-valenced associations. Importantly, this interpretation may provide insight into why, in Study 1, we did not observe a prevalence of negative triads: individuals experiencing suicidal ideation may use affective compartmentalization to orient their mental associations toward positively-valenced contents in general, while concomitantly reserving negatively-valenced contents for their perceptions of themselves.
Thus, negative concepts, being related primarily to the self, do not participate in many triangles in the network and this is able to explain why wee see a larger amount of $\{-,-,-\}$ triplets only in the label-reshuffled null model (see~Fig.~\ref{sb:co}).

\section*{Study 3: Understanding semantic frames and emotional perceptions in suicide notes}

After having investigated structural balance and self-perceptions, we turn the attention to the overall layout of conceptual and emotional perceptions in the CO network and compare these against the baseline linguistic model provided by free associations.

\subsection*{Conceptual relevance measured via network metrics}

In cognitive network science, conceptual distance successfully identifies conceptual relatedness~\cite{gray2019forward,kenett2017semantic,vitevitch2019network} (see also the SI: Semantic network distance and word prominence). The closeness ranks reported in Table~1~(SI) indicate the most prominent concepts identified in suicide notes through co-occurrence associations. Non-meaningful words such as determiners, adpositions and conjuncts were removed from the ranking as they do not convey any meaning in isolation. \rev{Among the top central words for closeness, we selected “love”, “want”, “help” and “life” as they were the most frequent words in the original corpus not being stopwords. We also focused on “I” in order to identify the semantic framing of the authors of suicide notes when they address themselves.}

\rev{``Love'' is the concept with the highest closeness centrality in the whole CO network.} This indicates that suicide notes featured ``love'' in a wide array of different contexts and corroborates the earlier results based on the SVO representation. Other prominent concepts in the CO network include moving verbs like ``go'', ``get'', ``make'' and ``take'' and other words expressing desire such as ``want'', ``live'' and ``help'', which are all interconnected. ``Live'' and ``life'' are prominently featured, together with specific aspects of life such as ``work'' and ``family''. These words indicate that suicide notes are not explicitly dominated by concepts directly related to suicide, as none of the 40 top-ranked concepts reported in Table~1~(SI) are directly related to suicide or death in general. Instead, these concepts portray rather general aspects of life, even though the triad ``want''/``live''/``help'' indicates a willingness to get better and a focus on the topic of life.

Notice that ``love'' is the only top-ranked verb/noun that would significantly drop by several ranks (6) in closeness rankings based on random configuration models. This means ``love'' is ranked higher than random expectation in the empirical suicide notes. Free associations limited to the same set of words in common with co-occurrences identified ``love'' as less central than concepts like ``time'' and ``money'' (see~SI:~Table~2). Hence, in the process of mind-wandering free from suicidal ideation, as captured by free associations, ``love'' is not as central as it is in suicide notes. These two comparisons provide compelling quantitative evidence that ``love'' is an exceptionally prominent concept for the authors of the notes.

The reconstruction of the semantic frame and emotions linked to ``love'' is thus relevant for the investigation of suicide notes. People who committed suicide might interpret or alter the commonly positive meaning attributed to ``love'' in mainstream language. This alteration takes place on both the semantic and emotional levels of language processing calling for additional analyses on these levels.

\subsection*{Emotional profiles of concepts in the CO and free association networks}

Semantic prominence is not enough, on its own, to understand how authors perceived love and other concepts in their final notes. To address this, we turn to sentiment analysis and emotional profiling. Sentiment features of suicide notes helped in the supervised detection of suicidal ideation in text identification tasks (cf.~\cite{schoene2016automatic}). In the following, emotional profiles are presented by: (i) considering the observed emotional richness of a word or set of words against random expectation; (ii) reporting the \textit{z}-score of the observed richness against random word sampling from the NRC Emotion Lexicon (see SI).

\begin{figure}[h!]
\centering
\includegraphics[width=0.61\textwidth]{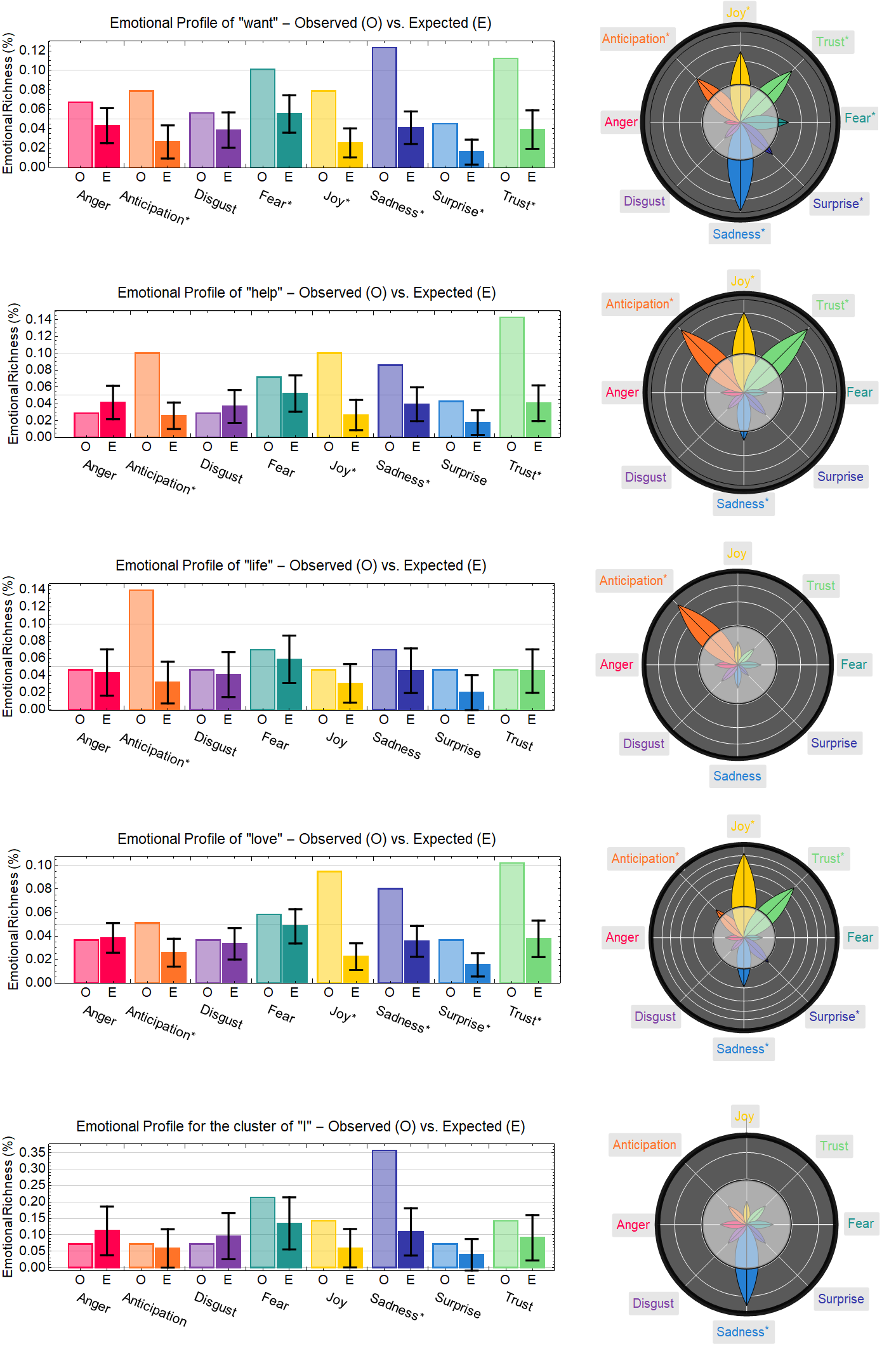}
\hspace{0.5cm}
\frame{\includegraphics[width=3.24cm]{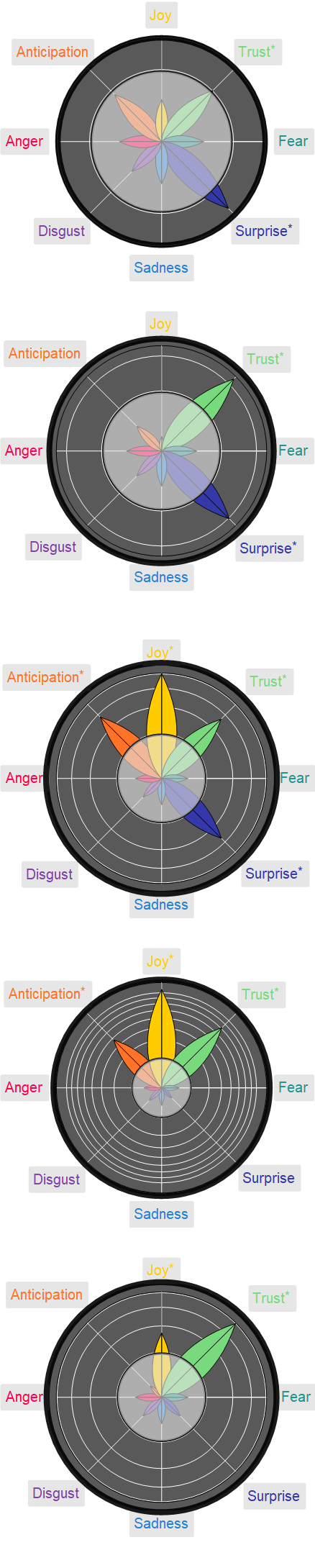}}
\caption{Reconstructed emotional profiles for ``want'', ``help'', ``life'', ``love'' and "I" in the co-occurrence network, visualized either as a comparison between observed and randomly expected emotional richness (left) or as an emotion wheel of \textit{z}-scores (right). Error bars indicate one standard deviation over a sample of 1000 random iterations. Emotions more frequent than random expectation, given a significance level of $\alpha = 0.05$, are marked with an asterisk (plots on the left) and fall outside of the gray circle (plots on the right). White circles in the emotional wheels count \textit{z}-scores, e.g. ``life'' elicited anticipation with a \textit{z}-score of around 4. The framed visualizations contain emotion profiles for the same words on the left but coming from the free association network and referring to mind-wandering in the absence of suicidal ideation.}
\label{fig:emoSC}
\end{figure}

Figure~\ref{fig:emoSC} reports the emotional profiles for highly central concepts identified in the CO network, namely ``love'', ``want'', ``help'', ``life'', \rev{and ``I''.} \rev{The semantic frames for all these concepts include emotions that are present in suicide notes but are rather atypical for how they appear in common language (as captured by mind-wandering through free associations). This is to say that these highly central concepts are all common words but their semantic frames in suicide letters are shaped by emotions that differ drastically from the ways of thinking among healthy people associating “love”, “want”, “help” and “life” with the first ideas that come to their minds.}

\subsection*{Emotions in suicide notes differ from common language}

What do people who committed suicide ``want''? This delicate question can be explored through the reconstructed semantic frame/emotional profile of ``want''. In the CO network such concept is linked with words associated with anticipation and positive emotions such as joy and trust, but also with ones eliciting negative emotions including fear and sadness. All these emotions but surprise are absent in the semantic frame of ``want'' coming from free associations. This difference indicates that the emotions of anticipation, joy, fear and sadness characterize ``want'' as a proxy for both positive and negative ideas in the context of suicidal ideation. The emotional profile and semantic frame of ``want'' suggest that what the authors of suicide notes express with ``want'' is not only the desire of positive, joyous or trustful things but also sad or frightening ideas. Differently from the common willingness to get better through positive experiences~\cite{plutchik2013theories}, the authors of suicide notes also express a willingness to get through sad or painful experiences, such as the fatal act of taking one's own life.

The concept of ``help'' is associated with words eliciting mostly positive emotions, featuring a high level of anticipation or projection into the future, e.g. looking for help with future events or plans. The trust associated to ``help'' in suicide notes is common with the linguistic baseline of free associations. In suicide notes, sadness is also featured more prominently than random expectation around ``help''. Notice that anticipation and sadness reveal evidence of resignation, according to Plutchik's atlas of emotions~\cite{plutchik2013theories}, an emotion absent in free mind-wandering.

``Life'' is a concept with positive connotations in the free association baseline model, where it is strongly linked to trust and joy. In the language of suicide notes, the semantic frame of ``life'' elicits no significantly positive emotions. \rev{It is worth underlining that suicide notes frame ``life'' as a trust-less concept whereas in mind-wandering it has mostly trustful associations. In suicide notes, only an increased level of anticipation was found, which in itself is a rather neutral projection into the future. Both these patterns are expected in the context of final requests extending after the end of one's life
}~\cite{pestian2010suicide,schneidman1981suicide,handelman2007content,foster2003suicide}. Besides anticipation, ``life'' in suicide notes is framed in a way devoid of any positive or negative emotions, a striking result that calls for additional discussion in view of psychological theories like narrative psychology.

``Love'' was found to be the most prominent concept featured in suicide notes. The semantic frame of ``love'' consists of words eliciting joy and trust. This indicates a positive perception of ``love'' itself in line with its positive perception in common language and found also in free mind-wandering, devoid of suicidal ideation, as represented here by free associations. \rev{Nonetheless, in suicide notes sadness was also present in the semantic frame of ``love'', as well as a reduced degree of anticipation, suggesting a more nuanced perception of this concept in comparison to free associations. Since anticipation indicates a projection of ideas into the future, the reduced anticipation and increased sadness found in suicide notes would indicate a melancholic framing of love, in contrast with the positive connotation found in mind-wandering and, thus, in common language.}

In Study 2 we identified an overall negative perception of self in suicide notes. \rev{The semantic frame reconstructed in the present study better characterizes the perception of ``I'': self-perception in suicide notes revolves around sadness and it crucially lacks trust. This is different from the linguistic baseline represented by free associations; in fact, in the absence of suicidal ideation people framed ``I'' along with positive emotions like trust and joy. Both of these emotions are absent in suicide notes. This highlights a negative, trust-less perception of the self portrayed by suicide authors and it provides further evidence of psych-ache, as reported in Study 1.}

The above results indicate that the meaning communicated through individual concepts is not the same as found in common language. Instead, suicide notes introduce richly nuanced contexts that shift the meaning and emotional content attributed to many concepts, making it fundamental to take these contexts into account for correctly understanding suicide notes, as pointed out by recent investigations, cf.~\cite{galasinski2017discourses}.




\section*{Discussion}

This first-of-its-kind study uses cognitive network science for identifying key concepts typical for suicide notes and reconstructing the meaning and emotions from the final words expressed by authors who committed suicide.

Structural balance is a potential mechanism that has been long theorized to drive certain aspects of cognitive organization, particularly with regard to resolving conflicting beliefs or attitudes~\cite{heider}. Pairing this with other psychological theories, such as \textit{narrative psychology}~\cite{mcadams2001psychology,josselson2015narrative} and the \textit{meaning-maintenance model}~\cite{heine2006meaning,proulx2006death}, we can consider how the patterns observed in suicide notes fit with a broader understanding of the psychological literature.

According to the meaning-maintenance model, people are fundamentally driven to construe their lives, perceptions, and behaviors as meaningful~\cite{heine2006meaning}. This is a position long held by existentialist philosophers, and is widely accepted by contemporary psychologists~\cite{yalom2020existential}. Moreover, given the drive to make meaning from our experiences and perceptions of the world~\cite{plutchik2013theories}, people will be motivated to restore the sense of meaningfulness whenever perceptions of their own life's meaning are threatened. Suicidal ideation might represent a response to such a threat, but also poses a challenge to identifying the meaning of one's life. In this perspective, writing a suicide note may represent a way of re-establishing a sense of meaningfulness and coherence in the face of circumstances that led the individual to consider or complete suicide.

This drive to find meaningfulness in narratives, notes, and letters is supported by narrative psychology, which focuses on the function, structure, and contents of the stories we tell ourselves and others about life~\cite{mcadams2001psychology}. With meaning-making as a distal motive for writing suicide notes, we may then interpret the structural balance and positive emotional perceptions reported above as concrete signals of meaning-making, or rather as proximal communicative mechanisms through which meaning-making can be more readily achieved.

In short, we argue that: (1) people are driven to perceive their lives as meaningful and coherent, (2) they use narratives or story-telling as a way to encapsulate and restore these perceptions when threatened, and that (3) potential ways of improving the coherence of one's own psychological narratives is by introducing balance and positive emotional semantic frames to otherwise unbalanced or negative sets of cognitions.

The relevance of this reasoning to the results of the present study can be summarized in two key elements. On the one hand, our extension of structural balance~\cite{heider} to networks of valenced conceptual associations provides quantitative evidence that the content of suicide notes tends to feature more positive triads than valence-reshuffled null models. This indicates that both the syntax and the valence of words used in suicide notes convey a tendency to avoid conflicting cognitions by assembling together pleasant concepts, in line with previous qualitative studies~\cite{o1999thematic} and quantitative investigations using a ``bag-of-words'' approach~\cite{handelman2007content}. On the other hand, semantic framing and emotional profiling both denoted suicide notes as being rich in positive/trustful perceptions revolving around concepts such as ``love'', ``take'', ``go'' and ``way''. These positive emotional portrayals also contained strong signals of sadness and anticipation of the future.

Notice that despite the overall prominence of positive, compartmentalized conceptual structures in the mindset of suicide notes, our analysis also found that self-perception is mostly dominated by negative associations. The semantic frame and the subject-verb-object associates of ``I'' highlighted negative semantic relationships of self and others, mostly dominated by sadness and absent in the linguistic baseline model provided by free associations (in absence of suicidal ideation). This represents compelling evidence for a \textit{cognitive dissonance} in the mindset of people who committed suicide. Suicide notes denote a high level of structural balance with positive conceptual triads but also a negative cluster of concepts surrounding the self but lacking triadic closure (as evident from the SVO analysis in Figure~\ref{fig:svo-main}).

Our results integrate and extend previous findings and also show ``love'' is a central concept in suicide notes. Recent studies debated whether the emotional perception of ``love'' in suicide notes is as positive as in common language~\cite{lester2016comparison,galasinski2017discourses}. Our network approach enriched with linguistic annotations identified ``love'' in suicide notes as being attributed to the same positive connotations as appear in common language, but also imbued with a sense of sadness. Furthermore, while love is central across suicide notes, it is described as mostly related towards other people in diverse ways, and does not function as a purely positive emotion (as assumed by~\cite{lester2016comparison}). This indicates the importance of going beyond considering words in isolation~\cite{handelman2007content} to better understand suicide notes. Our quantitative approach reconstructs words as interconnected in the language of suicide notes and compares it against random network models and linguistic baseline models (free associations~\cite{de2019small}). This network structure reveals ``love'' as: (i) being prominent in the considered narratives, even more than in mind-wandering as captured by free associations, (ii) being focused on relations with others and (iii) eliciting a nuanced set of emotions consisting of joy and trust (as in free associations) but also nuances of anticipation and sadness. Structuring narratives around trustful and joyous relationships with loved ones aligns well with the above interpretation of suicide notes as being strategically driven by meaning identification. Another outcome of such strategy, aimed at avoiding conflicting cognitions, might be the signals of anticipation and sadness attributed to ``love'', which both identify resignation, a passive acceptance of threats that generates no anger or conflict at the cost of feeling defeated and incapable of creating change. This perception calls for future research investigating the psychological mechanisms at work. \rev{However, our approach already suggests a way in which emotional valences in text data such as suicide notes can be measured in a more contextualized manner by combining general population-level estimates with case-specific structure of associations between different concepts/words. Thus it can potentially be used more generally for estimating sentiment associated with different words while accounting for the specificity of a given corpus, a problem already recognized by several authors (cf.~\cite{sienkiewicz2016impact}).}

All in all, the reconstructed contexts of concepts in suicide notes provide evidence for meaning-making narratives, aimed at coping with threats through meaning identification and conflict-avoidant storytelling. This rich landscape is invisible when considering words in isolation and only emerges from the complex structure of conceptual and emotional connections between words in text. Cognitive network science~\cite{kenett2017semantic,castro2020contributions,vitevitch2019network,stella2020text}, combining psycholinguistics, computer science and network science, represents a powerful framework for reconstructing conceptual relationships, opening a window into people's minds, along with their subjective perceptions and perspectives. The ability for cognitive network analyses to parse large volumes of texts without human supervision calls for future large-scale investigations of the cognitions and perceptions expressed in suicide notes.



\subsection*{Limitations and Future Directions}

This study does not use machine learning for automatic classification of texts. It rather focuses on the reconstruction of the general mindset expressed through genuine suicide notes which represents typical streams of thoughts of people committing suicide. Achieving this quantitative knowledge is key not only for better understanding of suicide notes but also for empowering interpretable future models of automatic language processing~\cite{schoene2016automatic,al2018absolute,pestian2012sentiment,rudin2019stop}.

The main limitation of the present study is the lack of comparison with a different set of texts. One might consider comparing suicide notes against other corpora, e.g. love letters~\cite{schoene2016automatic}. However, these comparisons would include potential issues with unexpected content as found in text, as the overall topic of a corpus offers little guarantee on its linguistic content and semantic frames~\cite{hassani2020text}. \rev{For instance, love letters might frame ``love'' in nuanced - even sad - ways because of lovestruck authors or thoughts related to regret and desire, while very similar language and mindsets can be found also in suicide} letters~\cite{o1999thematic,schneidman1981suicide}. \rev{Such emotional/semantic overlap may lead to considerable errors when comparing different corpora, making it difficult to produce interpretable results. In this work, we still managed to compare suicide notes against a linguistic baseline model by adopting free} associations~\cite{de2019small}, \rev{which come from mind-wandering and are devoid of suicidal ideation or any other systematic frame of reference by construction. Since many of the emotional and structural differences we found between suicide notes and free associations are supported by relevant literature in clinical psychology, it suggests that our methodological framework is suitable for further analysis of autobiographical corpora, which we will address in future research.}

Concerning the \rev{emotional} balance of these cognitive networks, the major limitation is the inability of observing the $\{-,+,+\}$ triads, which creates a weak bias towards balanced triangles. This was \rev{only} a minor issue because we were able to observe that even with this constraint, depending on the null model, different levels of balance were observed. Moreover, null models were subject to the same limitation, and so still provided appropriate comparisons within the context. A potential extension of this analysis could be based on the investigation of how to define neutral links between words. \rev{Future research might also investigate more elaborate definitions of emotional balance, based not only on valence but also on other dimensions of affect like arousal or dominance. Based on recent results in the relevant literature} \cite{gorski2020homophily}, \rev{adding multiple variables for defining balanced and unbalanced triads might reduce the observed degree of balance.}

Another limitation is the identification of emotional profiles in suicide notes based on cognitive datasets referring to everyday language usage~\cite{mohammad2013crowdsourcing}. This problem is partially addressed by reconstructing emotional profiles from the specific syntactic relationships detected in suicide notes. In this way, attention should be given not to the emotions of individual words but rather to the way such words are interconnected within the observed texts. Building and adopting emotional lexical resources extracted from texts with suicidal ideation could provide more accurate readings of emotional profiles, and so represent important goals for further research.

\rev{In our approach, we operationalize the affective information of each word via norms obtained from large-scale studies grounded in affect theory \cite{warriner2013norms}. It is important to note, however, that past studies have given reasons to be cautious about the fidelity of automatic operationalizations in sentiment analysis, especially due to issues of individual variability, cultural influences and training biases \cite{cambria2017sentiment}. We note this concern, and highlight that our approach attempts to improve on this issue by reconstructing the variability of word usage in terms of network associations.}

\rev{
Furthermore, the present analysis is based on a corpus of a limited size and scope, both spatiotemporally and culturally. This is largely due to the limited accessibility of datasets of genuine suicide notes. Hence, our results should not be generalized without careful consideration, in particular concerning the cultural specificity of the sample. It also means that intercultural comparisons of language and mindsets expressed in suicide notes may be an interesting avenue for future research. Crucially, the approach we propose in this paper could be used to extract knowledge and cognitive structures expressed through suicide notes which vary across different populations and thus could be used to inform construction of unbiased early detection methods and systems.
}

\rev{Last but not least, our analysis used only ,,offline'' letters while there is evidence
that language used in online communication is often significantly
different~\cite{chmielCollectiveEmotionsOnline2011}. Therefore,
our results should not be generalized without additional care to the context
of online communication.
At the same time, the general approach we propose could be in principle used to identify
most prevalent differences between languages and mindsets of offline and online suicide notes.}

\section*{Conclusions}

In this research we present the first application of network science to the quantitative analysis of genuine suicide notes. Our approach combines some of the unique advantages of automated text analysis with networks, while using theoretical tools from psychology (e.g., structural balance theory), to gain a more detailed understanding of underlying psychological states associated with suicide notes. Cognitive network methods allow us to move beyond comparatively opaque, ``black-box'' models for classifying suicide notes, as they extract key ideas and emotions embedded in text. This knowledge extraction allows researchers to address questions about higher-level psychological processes and test hypotheses based on theory. Although the present study was data-driven and therefore exploratory, it demonstrates that cognitive networks are a valid approach for future confirmatory investigations, leading to a greater understanding of new possibilities for suicide prevention.

\section*{Materials and Methods}

\subsection*{Dataset of suicide notes}

A genuine suicide note is a text left by a person who subsequently committed suicide. This investigation used the Genuine Suicide Notes corpus by Schoene and Dethlefs~\cite{schoene2016automatic}. The corpus represents a collection of 139 genuine suicide notes collected from fact-checked newspaper articles and other previous small-scale investigations of suicide notes. All notes are in English and were anonymized by changing names of people and places or any reference to identifying information. Shorter suicide notes, including those less than two sentences, were discarded. \rev{As mentioned above, this collection of suicide notes comes from sources like newspapers, books and diaries collected by clinical psychologists mostly in the US and in Europe. Notes were written and collected over a time window spanning over 60 years, between 1958 and 2016. The corpus was assembled by Schoene and Dethlefs} \cite{schoene2016automatic}. \rev{On average, a letter included 120 $\pm$ 12 words and a total of 2075 different concepts were stated in the whole corpus. Deleting stopwords, i.e. words not possessing a meaning when isolated from other concepts, led to 1909 different concepts being stated in the considered suicide letters.}

\subsection*{Constructing two types of linguistic networks}\label{sec:net}

This work implemented two types of network construction. Co-occurrence (CO) networks captured syntactic relationships between adjacent words in a sentence. Subject-verb-object (SVO) networks captured triplets of syntactic relationships between a subject, a verb and an object. These network representations of the structure of knowledge in suicide nodes was also enriched with sentiment labels (e.g. a word being perceived as positive/negative/neutral in common language, see~\cite{warriner2013norms}) and emotional labels (i.e. words eliciting one or more emotional states, see~\cite{mohammad2013crowdsourcing}). Additional details,
\rev{including a list of most common associations between concepts in the SVO networks
and a set of examples of sentences including some of those associations},
can be found in Supplementary Information (SI).

\subsection*{\rev{Emotional balance and triadic closure} analysis}

Sentiment labels can be positive, negative or neutral and are used for structural balance analysis. Let $G=(V,E)$ be an undirected and signed network, with $|V|$ vertices (words) and $|E|$ edges (co-occurrence relations). We define edge labels $w \in \{-1,1,0\}$ between two words (a,b) as follows: $w(a,b) = w(b,a) = 0$ if both words are neutral; $w(a,b) = w(b,a) = -1$ if either $a$ or $b$ is negative; and $w(a,b) = w(b,a) = 1$ if both $a$ and $b$ are positive or if one of them is positive and the other neutral. As a result, we obtain a signed network with 1962 positive links and 1362 negative links. We consider that the neutral links (links with label 0) do not play a role when calculating the degree of balance. 
With the above definition the shadowed triad from Figure~\ref{fig:sb} (unbalanced) is never observed in this signed network. The degree of balanced is obtained by calculating the fraction of balanced triads. \rev{Results were tested against configuration models \cite{viger2005fast}, which randomly rewires links while keeping the graph connected and fixing the degree of each word in the empirical network. Fixing the degree of words was key in accounting for local influences of degree over non-triadic patterns like preferential attachment \cite{hills2010associative,hills2009longitudinal} and degree assortativity \cite{steyvers2005large}}.

\subsection*{Free associations as a linguistic baseline model}

In the manuscript we build and investigate cognitive networks of conceptual associations representing how authors conceptually framed and perceived ideas in their last letters. These network structures have to \rev{be} compared not only against baseline network null models (e.g. configuration models in \rev{emotional} balance) but also against other linguistic baseline models, indicating how people in non-suicidal populations would have framed and interconnected the same set of concepts occurring in suicide notes. Whereas other investigations based on machine learning adopted textual corpora like love letters or blog posts about depression~\cite{schoene2016automatic} as linguistic baseline models, it is not clear what type of content or semantic frames should be expected from a given text corpus. For instance, love letters might present feelings of melancholia closely related to suicidal ideation, without clear or direct control from the experimenter. This uncertainty makes the comparison more difficult. For this reason, we followed another approach, using not texts but directly complex networks as baseline linguistic models. We focused on mind-wandering, a cognitive phenomenon where concepts are interconnected with each other in ways relying on memory only and free of additional constraints (e.g. linking only concepts related syntactically within a sentence). Mind-wandering is captured by free associations~\cite{de2019small,kenett2017semantic}, i.e. naming the first words coming up to mind when thinking of a certain concept. Hence, we use networks of free associations as baseline linguistic models representative of the mind-wandering of a large population of individuals without suicidal tendencies. Using the Small World of Words dataset~\cite{de2019small}, we built
a network of free associations. It contains 1581 concepts, all included in the original co-occurrence (CO) network, which are connected according to free/mind-wandering associations. These associative structure is then used in the main text as a linguistic baseline for investigating semantic frames and emotional perceptions found in suicide notes.


\section*{Acknowledgements}
The authors acknowledge the Winter Workshop on Complex Systems series, Cynthia S. Q. Siew, Benjamin Ortiz Ulloa and Narges Chinichian for valuable discussion. This work was supported by FCT, Fundação para a Ciência e a Tecnologia, under project UIDB/50021/2020.

\section*{Supporting Information}

\subsection*{Network Construction}

Two types of networks were adopted for the current analysis:
\begin{itemize}
    \item Co-occurrence (CO) network in which nodes represent concepts and links indicate succession relationships. Thus, relationships were captured through a sequential chain, establishing links between preceding and subsequent words. For instance, the sentence ``The pen is red'' would feature links ``the--pen'', ``pen--is'' and ``is--red''. This straightforward approach is reminiscent of word co-occurrence and can capture syntactic relationships in language, e.g. a word specifying semantic features of another one.

    \item Undirected, weighted network of relationships induced by SVO triplets extracted from the notes. Each SVO triplet was decomposed into three possible links included in a triple:
    \begin{enumerate}
        \item subject--verb
        \item verb--object
        \item subject--object
    \end{enumerate}
    For instance, the triple ``he--look--at'' consists of the following three pairs: (1) ``he--look'', (2) ``look--at'', and (3) ``he--at''. Edge weights are equal to the number of co-occurrences of two words in the same SVO triplets. Note that this way of decomposing SVO triplets into node pairs does not introduce any structural bias, as each component (subject, verb and object) appears exactly two times over three pairs generated from a single triplet. On the other hand, tokens playing central syntactic roles (i.e. subject and verbs) will appear in more SVO triplets than less central ones. Therefore, these tokens will have higher degrees and strengths (sums of edge weights) in an SVO network. This is an important property of our method which encodes some of the crucial semantic features of text corpora directly into degree/node strength distributions.

    In order to filter out errors and relationships between words that were accidental and/or idiosyncratic for individual suicide notes, the SVO network was limited only to relations (edges) with weights equal to or greater than $2$. In other words, only relationships that occurred at least twice over the entire corpus were considered. Then, all positive weights were decreased by $1$ to remove the truncation of the lower tail at $2$ in order to make sampling from the appropriate canonical ensemble (undirected weighted configuration model~\cite{squartini_unbiased_2015}) more feasible.
\end{itemize}

The above linguistic networks were also enriched with:
\begin{itemize}
    \item sentiment labels, i.e. positive/negative/neutral, indicating the overall pleasantness of a concept as expressed by large audiences in psycholinguistic studies about common language. These labels identified words in the lower quartile (negative), middle quartiles (neutral) and upper quartile (positive) of valence scores as obtained from~\cite{warriner2013norms};
    \item emotional labels, expressing the emotions elicited by a given word as indicated by large audiences in psycholinguistic studies about common language. The considered emotions were those from the NRC Emotion Lexicon~\cite{mohammad2013crowdsourcing}, namely anger, anticipation, fear, disgust, joy, sadness, surprise and trust.
\end{itemize}

Both the above sentiment and emotional datasets are relative to overall, global perceptions of concepts as represented in mainstream populations. Hence, these datasets are not directly informative about the subjective emotional perceptions portrayed by people who committed suicide. In order to reconstruct these subjective perceptions, network measures are required.

The CO and SVO networks enable the reconstruction of a semantic frame in terms of a network neighborhood of a given concept. As an example, the semantic frame of ``love'' is represented by the first neighbors of concepts syntactically related to ``love'' by people who committed suicide in their suicide notes. Checking this semantic frame/network neighborhood provides crucial information about the contexts and perspectives that featured ``love'' in the suicide notes. On the emotional level, sentiment and emotional attributes are not universal, as they could change according to the subjective perceptions of the authors or the context in which concepts appear. For instance, ``love'' is usually indicated as a positive word bringing emotions of trust and joy. But what about associations like
``betraying love'' or ``missing love''? Placing ``love'' in different contexts can alter its subjective perception. This is the main reason why language cannot simply be considered as a bag of words, i.e. a collection of isolated concepts, but rather as a network of interconnected linguistic units whose meaning and emotions can change according to the way they are networked together. Therefore the above networks provide access to the associative and emotional perceptions of conceptual entities in the minds of people committing suicide.

\subsection*{Additional natural language processing}\label{sec:svos}

Words in all notes were lemmatized and annotated with part-of-speech and dependency tags
as classified with a state-of-the-art NLP library
\textit{Spacy} (https://spacy.io)
based on OntoNotes (v5) annotated corpus and Penn Treebank~\cite{weischedel2013ontonotes}.
The tags specify roles played by particular words in a sentence as well as syntactic dependency relationships between them. This additional information was used to derive network representations of the notes capturing more fundamental syntactic links instead of simpler, sequential relationships between preceding and subsequent words. Specifically, it allowed a decomposition of all sentences into a kind of generalized subject-verb-object (SVO) triplets. Here, an SVO triplet consists of a subject, which is seen as an active agent, a verb seen as an action performed by the subject, and an object standing for anything that the action (verb) performed by the subject relates to. This is why the decomposition used here is somewhat more general, as any token other than nominal subject and subordinate to a verb in a syntactic tree of a sentence is considered an object.

For instance, in our approach the following sentence is decomposed into four different
SVO triplets:

\begin{itemize}
    \item He was looking at a tree, which was very tall.
    \begin{enumerate}
        \item he--look--at
        \item he--look--tree
        \item tree--be--very
        \item tree--be--tall
    \end{enumerate}
\end{itemize}

As the example shows, the method disaggregates relative clauses such as ``tree, which is tall'' into separate SVO triplets. This way it is more capable of capturing the semantics of compound and complex sentences. Moreover, more important meaning making tokens (such as ``he'', ``look'' and ``tree'' in the example above) appear in multiple triplets by design. Because of this, even simple summaries such as frequencies of words over all SVO triplets can capture important semantic features of a text corpora.

The second important processing step was a custom lemmatization which accounted for the specific way in which the suicide notes were anonymized. Most words were lemmatized according to the standard rules implemented for the English language in \textit{Spacy}. However, all names of persons in the notes were substituted with several generic placeholder names, such as Jane or William, and so were reduced to a special ``s/he'' lemma. Moreover, all occurrences of ``he'' and ``she'' were also lemmatized this way.

\subsubsection*{Extracting SVO triplets}
The general procedure used for extracting SVO triplets is relatively straightforward. First, documents are tokenized into sentences and sentences are tokenized into words.
A word is considered \textbf{semantic} if it is a (proper) noun, pronoun, verb, adverb, adjective, adposition or a negation modifier (not). Then, all semantic words are assigned with one of the following classes:
\begin{itemize}
    \item \texttt{SUBJECT.} Words which are either active or passive nominal or clausal subjects (according to syntactic dependency tags).
    \item \texttt{VERB.} Words which are either verbs (according to part-of-speech tags) or relative clauses (according to syntactic dependency tags).
    \item \texttt{OBJECT.} Any semantic word which is neither a verb nor a subject.
\end{itemize}

Additionally, we define two procedures.

\begin{itemize}
    \item \texttt{GET\_VERB(token).} \\
    Return nearest \texttt{VERB} word which is above \texttt{token} in the syntactic dependency tree.
    \item \texttt{GET\_SUBJECT(token).} \\
    Define \texttt{verb = GET\_VERB(token)}. \\
    Define \texttt{subj} to be the nearest \texttt{SUBJECT} word below \texttt{verb} in the syntactic dependency tree. \\
    If \texttt{subj} is a (proper) noun according to its POS tag, return \texttt{subj}. \\
    Otherwise, return the nearest \texttt{NOUN} word above \texttt{subj} in the syntactic dependency tree.
\end{itemize}

Finally, SVO triplets are extracted from a sentence according to the following procedure:
\begin{itemize}
    \item For each \texttt{word} in the sentence:
    \begin{itemize}
        \item If \texttt{word} is \texttt{OBJECT}:
        \begin{itemize}
            \item Return a triplet:  \texttt{GET\_SUBJECT(word)}, \texttt{GET\_VERB(word)}, \texttt{word}.
        \end{itemize}
    \end{itemize}
\end{itemize}

For instance, in the example sentence ``He was looking at a tree, which was very tall'' there are four \texttt{OBJECT} words which are mapped to four SVO triplets:
$$
\begin{array}{lcclcl}
    1. & \texttt{at} & \mapsto & \texttt{GET\_SUBJECT(at), GET\_VERB(at), at} & \mapsto & \texttt{he, look, at}
    \\
    2. & \texttt{tree} & \mapsto & \texttt{GET\_SUBJECT(tree), GET\_VERB(tree), tree} & \mapsto & \texttt{he, look, tree}
    \\
    3. & \texttt{very} & \mapsto & \texttt{GET\_SUBJECT(very), GET\_VERB(very), very} & \mapsto & \texttt{tree, be, very}
    \\
    4. & \texttt{tall} & \mapsto & \texttt{GET\_SUBJECT(tall), GET\_VERB(tall), tall} & \mapsto & \texttt{tree, be, tall}
\end{array}
$$

\subsection*{Structural balance theory}
Structural balance theory, first explored by Heider~\cite{heider}, states that for a signed triad to be balanced the product of its signs must be positive. Thus, from the four possible triads -- $\{+,+,+\}$, $\{-,+,+\}$, $\{-,-,+\}$, $\{-,-,-\}$ -- only the first and third are considered balanced. As an example, if we think about the following statements ``a friend of my friend is my friend'', ``an enemy of my enemy is my friend'', along with similar others, we are able to verify that they follow the concept of balance as defined by Heider.

Some years later, Cartwright and Harary generalized the concept of structural balance to social networks, introducing signed graphs in which edges had positive and negative signs corresponding to positive or negative ties between individuals~\cite{harary1953,cartwright1956structural}. They extended the concept of balanced triads to balanced networks by allowing cycles with more than three edges. A cycle is considered balanced if the product of the signs of its edges is positive, i.e., if there are no odd number of negative edges in a cycle. To measure structural balance, Harary introduced the concept of \textit{Degree of Balance} (DoB) of a signed network as the ratio of the number of positive cycles to the total number of cycles
~\cite{harary1959measurement}. Let $G$ be a signed graph, $c(G)$ be the number of cycles of $G$, $c_{+}(G)$ be the number of positive cycles of $G$, and $DoB(G)$ be the degree of balance of $G$. Then:
\begin{equation}
    DoB(G) = \frac{c_{+}(G)}{c(G)}.
\end{equation}
In this work we use cycles of size three -- triads.

Even though these theories were developed more than half a century ago, it has only been in the last decade that structural balance theory has been revivified in different domains. Recently, Chiang et.al ~\cite{chiang2020triadic} presented a study exploring triadic balance in the brain regarding how brain activity expresses possible cognitive (im)balances when people are faced with cooperative decisions regarding social dilemmas. They showed that an individual's psychological states are reflected in the different areas of the brain that are activated when they are situated in unbalanced or balanced triads. When encountering unbalanced triads, individuals showed activation in brain regions associated with cognitive dissonance, reinforcing Heider's theory.

\subsection*{Semantic network distance and word prominence}

The identification of semantically related concepts is traditionally performed by semantic latent analysis, which maps the problem of measuring conceptual distance into selecting appropriate metrics in a vectorial space of words. However, in predicting semantic relatedness, semantic latent analysis was recently outperformed by network distance in cognitive networks, i.e. counting the smallest number of conceptual links connecting any two concepts in the same connected component of a given network~\cite{kenett2017semantic}. We build upon this evidence and define conceptual relatedness as concepts being at a shorter network distance. A concept which is related to, i.e. at shorter network distance from, almost all other concepts must be prominent. This intuitive definition of conceptual prominence is methodologically implemented by closeness centrality~\cite{newman2018networks}, which identifies concepts that are at shorter network distance from all other connected concepts.

Closeness centrality has been successfully used in previous investigations as a proxy of conceptual prominence predicting word acquisition (cf.~\cite{stella2017multiplex}). Also in the current analysis, we use closeness centrality as a quantitative way for identifying prominent concepts in the reconstructed mindset around suicidal ideation.

As a statistical baseline, the closeness centrality of concepts in the empirical networks was matched against closeness centrality in configuration models~\cite{newman2018networks}, i.e. random networks fixing the empirical degrees of words but otherwise randomizing conceptual links.

\begin{table}[ht]
    \centering
    \begin{tabular}{cccc}
    \text{Rank} & \text{Co-occ. Netw.} & \text{Free Asso.} & \text{Co-occ. Restr.} \\
     1 & \text{be} & \text{time} & \text{be} \\
     2 & \text{have} & \text{money} & \text{have} \\
     3 & \text{to} & \text{love} & \text{to} \\
     4 & \text{and} & \text{work} & \text{and} \\
     5 & \text{of} & \text{good} & \text{of} \\
     6 & \text{love} & \text{sad} & \text{love} \\
     7 & \text{for} & \text{food} & \text{do} \\
     8 & \text{do} & \text{home} & \text{for} \\
     9 & \text{in} & \text{day} & \text{in} \\
     10 & \text{take} & \text{life} & \text{take} \\
     11 & \text{go} & \text{happy} & \text{go} \\
     12 & \text{get} & \text{out} & \text{get} \\
     13 & \text{way} & \text{death} & \text{make} \\
     14 & \text{one} & \text{place} & \text{live} \\
     15 & \text{make} & \text{water} & \text{way} \\
     16 & \text{live} & \text{child} & \text{one} \\
     17 & \text{help} & \text{help} & \text{help} \\
     18 & \text{want} & \text{person} & \text{want} \\
     19 & \text{time} & \text{fun} & \text{time} \\
     20 & \text{day} & \text{bad} & \text{work} \\
     21 & \text{call} & \text{letter} & \text{day} \\
     22 & \text{end} & \text{me} & \text{end} \\
     23 & \text{life} & \text{car} & \text{all} \\
    24 & \text{thing} & \text{man} & \text{life} \\
    25 & \text{work} & \text{blue} & \text{call} \\
    26 & \text{all} & \text{house} & \text{not} \\
    27 & \text{not} & \text{red} & \text{feel} \\
    28 & \text{feel} & \text{now} & \text{thing} \\
    29 & \text{family} & \text{book} & \text{this} \\
     30 & \text{give} & \text{save} & \text{give} \\
    \end{tabular}
    \caption{Top 30 concepts ranked in descending order based on their closeness centrality in the original co-occurrence network (Co-occ. Netw.), the baseline network of free associations (Free Asso.) and the original co-occurrence network restricted to the same concepts featured in the free association network (Co-occ. Restr.).}
    \label{tab:freeassocent}
\end{table}

SI Table \ref{tab:freeassocent} reports the most prominent concepts, based on closeness centrality, in the original network of co-occurrences (with 2075 words) and the baseline free association network (with 1577 words, a subset of words in the co-occurrence network). As an additional check, words are ranked also in the subgraph of the co-occurrence network featuring only those words present in the free association network. Co-occurrences in suicide notes reflect a structural organisation of knowledge where love is more central than other topics, a pattern that is not observed in the baseline free association dataset, which features "time" and "money" as concepts with a higher closeness to all other connected words. Even by performing node alignment between the co-occurrence and the free associations networks, i.e. considering a subgraph of co-occurrences only between words present in the free association network, "love" remains more central than other concepts. These results indicate that "love" in the organisation of knowledge as assembled by authors of suicide notes was more central than expected in the knowledge of mindwandering as represented by free associations.

\subsection*{Network degeneracy}
Degeneracy of a network~\cite{klein_emergence_2020} measures the tendency that a random walker starting from a random node after one step ends up in a limited set of central nodes. Let $\mathbf{W}$ be a normalized weighted and undirected adjacency matrix such that weights in each row $\mathbf{W}_{i\bullet}$ sum up to $1$ so they can be interpreted as probability distributions over the next position of a random walker starting at node $i$. Then, degeneracy of the graph represented by $\mathbf{W}$ is defined as a normalized difference between maximal and observed entropy of the probability distribution $\mathbf{W}_{i\bullet}$ averaged over all nodes $i = 1, \ldots, N$ assuming they are equally likely:
\begin{equation}\label{eq:degeneracy}
    degeneracy =
        \frac{\log_2{N} - H(\langle\mathbf{W}_{i\bullet}\rangle)}{\log_2{N} - S} \in [0, 1]
\end{equation}
where $\langle\mathbf{W}_{i\bullet}\rangle$ is the vector of column means of $\mathbf{W}$, $\langle\mathbf{W}_{i\bullet}\rangle = 1/N\sum_{i=1}^N \mathbf{W}_{i\bullet}$, and $H(\cdot)$ is Shannon entropy. The $S$ term is $H(\langle\mathbf{W}_{i\bullet}\rangle)$ of an undirected star graph with $N$ nodes, that is:
\begin{equation}\label{eq:degeneracy-star}
    S = -\frac{N-1}{N}\log_2\left(\frac{N-1}{N}\right) - \frac{1}{N}\log_2\left(\frac{1}{N}\right)
\end{equation}

\subsection*{Emotional Profiling}

Emotional profiling was performed by labeling words according to the emotion they elicit, as indicated in the NRC emotion lexicon~\cite{mohammad2013crowdsourcing}. The dataset included 8 basic emotional states, whose combination can give rise to a wide variety of nuanced emotion. Emotional profiling was performed as in previous studies~\cite{stella2020text}, considering the number of words eliciting a given emotion in a certain network region, e.g. in the network neighborhood of a certain concept. The emotional profile of a word $p$ was considered by counting the fraction $f_i(p)$ of words syntactically linked to $w$ eliciting emotion $i$. By definition $f_i(p)$ ranges between 0 (no conceptual associates of $p$ elicit emotion $i$) to 1 (all concepts linked to $p$ elicit emotion $i$). As a reference model, we used a random sampling of words fixing the empirical sample size, e.g. the number of words syntactically linked to $p$, but neglecting empirical syntactic associations. Counting the fraction of randomly sampled words eliciting emotion $i$ provided direct-sampling distributions of random emotional profiles. We used these random distributions in order to attribute a \textit{z}-score to the observed emotional profile $f_i(p)$. This statistical procedure, with a significance level fixed at $\alpha = 0.05$, enabled a comparison of the strength of emotions elicited by individual concepts in our considered networks.

On the one hand, the visualization of \textit{z}-scores facilitates the immediate understanding of which were the stronger emotional intensities elicited by a given concept. On the other hand, the comparison provides additional information about how rich a concept can be in associations eliciting a given emotion.


\begin{table}[h!]
\small
\renewcommand{\arraystretch}{.8}
\centering
\begin{sffamily}
\begin{tabular}{c|cccc|c|cccc}
  \hline
 Rank & Word & \textit{z}-score & Significant? & Rank drop & Rank & Word & \textit{z}-score & Significant? & Rank drop \\
  \hline
    1 & love & 2.065 & Yes &   6 &  21 & feel & 0.389 &  & 6 \\
    2 & do & 1.134 &  &   2 &  22 & family & 1.220 &  & 116 \\
    3 & take & 0.884 &  &   6 &  23 & give & 0.113 &  &   2 \\
    4 & go & 1.027 &  &   6 &  24 & tell & 0.222 &  &  -3 \\
    5 & get & 0.780 &  &   4 &  25 & try & 0.733 &  &  17 \\
    6 & way & 1.328 &  &  20 &  26 & know & 0.318 &  &  -6 \\
    7 & one & 1.424 &  &  33 &  27 & start & 1.408 &  &  55 \\
    8 & make & 0.691 &  &   8 &  28 & good & 0.988 &  &  33 \\
    9 & live & 1.197 &  &  30 &  29 & as & 0.193 &  &  -3 \\
   10 & help & 1.261 &  &  20 &  30 & friend & 1.174 &  &  82 \\
   11 & want & 0.782 &  &   9 &  31 & year & 0.865 &  &  43 \\
   12 & time & 1.339 &  &  33 &  32 & money & 0.997 &  &  45 \\
   13 & day & 1.217 &  &  54 &  33 & leave & 0.530 &  &  13 \\
   14 & call & 1.016 &  &  16 &  34 & hope & 0.797 &  &  30 \\
   15 & end & 1.278 &  &  53 &  35 & child & 1.281 &  & 121 \\
   16 & life & 1.259 &  &  61 &  36 & think & 0.648 &  &  34 \\
   17 & thing & 1.224 &  &  28 &  37 & use & 1.259 &  & 102 \\
   18 & work & 1.274 &  &  38 &  38 & other & 1.077 &  & 146 \\
   19 & all & 0.306 &  &   4 &  39 & mother & 1.205 &  & 129 \\
   20 & not & 3.102 & Yes & -13 &  40 & come & 0.244 &  &  -5 \\
   \hline
\end{tabular}
\end{sffamily}
\caption{Top 40 concepts based on closeness centrality in the CO network. To detect how empirical syntactic relationships contributed to closeness centrality, 1000 configuration models with random relationships were built. Empirical concept centrality was compared against mean random expectation, enabling a measurement of the mean rank drop due to randomizing conceptual links. ``Love'' was found to be 6 positions higher in the empirical ranking than in random configuration models and was the only one among the top words for which rank drop was statistically significant.}
\label{tab:1}
\end{table}

\bibliography{main}

\end{document}